\documentclass[fleqn,usenatbib]{mnras}

\usepackage{newtxtext,newtxmath}
\usepackage[T1]{fontenc}
\usepackage{ae,aecompl}

\usepackage{graphicx}	% Including figure files
\usepackage{amsmath}	% Advanced maths commands
\usepackage{amssymb}	% Extra maths symbols

\newcommand{\beq}{\begin{equation}}
\newcommand{\eeq}{\end{equation}}
\newcommand{\bea}{\begin{eqnarray}}
\newcommand{\eea}{\end{eqnarray}}
\newcommand{\gae}{\lower 2pt \hbox{$\, \buildrel {\scriptstyle >}\over {\scriptstyle
\sim}\,$}} 
\newcommand{\lae}{\lower 2pt \hbox{$\, \buildrel {\scriptstyle <}\over {\scriptstyle
\sim}\,$}}
\title[Marginally fast synchrotron]{Marginally fast cooling synchrotron models for prompt GRBs}

\author[Beniamini, Barniol Duran $\&$ Giannios]{Paz Beniamini$^{1}$\thanks{Email: paz.beniamini@gmail.com}, Rodolfo Barniol Duran$^{2}$, Dimitrios Giannios$^{3}$ \\
	$^{1}$Department of Physics, The George Washington University, Washington, DC 20052, USA \\	
	$^{2}$Department of Physics and Astronomy, California State University, Sacramento, 6000 J Street, Sacramento, CA 95819, USA\\
	$^{3}$Department of Physics and Astronomy, Purdue University, 525 Northwestern Avenue, West Lafayette, IN 47907, USA
	}

\date{Accepted XXX. Received YYY; in original form ZZZ}

\pubyear{2018}

\begin{document}
\label{firstpage}
\pagerange{\pageref{firstpage}--\pageref{lastpage}}
\maketitle

\begin{abstract}
	Previous studies have considered synchrotron as the emission mechanism for prompt Gamma-Ray Bursts (GRBs). These works have shown that the electrons must cool on a timescale comparable to the dynamic time at the source in order to satisfy spectral constraints while maintaining high radiative efficiency. We focus on conditions where synchrotron cooling is balanced by a continuous source of heating, and in which these constraints are naturally satisfied. Assuming that a majority of the electrons in the emitting region are contributing to the observed peak, we find that the energy per electron has to be $E\gtrsim 20$ GeV and that the Lorentz factor of the emitting material has to be very large $10^3\lesssim \Gamma_{\rm em} \lesssim 10^4$, well in excess of the bulk Lorentz factor of the jet inferred from GRB afterglows. A number of independent constraints then indicate that the emitters must be moving relativistically, with $\Gamma'\approx 10$, relative to the bulk frame of the jet and that the jet must be highly magnetized upstream of the emission region, $\sigma_{\rm up}\gtrsim 30$. The emission radius is also strongly constrained in this model to $R\gtrsim 10^{16}$cm. These values are consistent with magnetic jet models where the dissipation is driven by magnetic reconnection that takes place far away from the base of the jet.
\end{abstract}

\begin{keywords}
radiation mechanisms: non-thermal -- methods: analytical -- gamma-ray bursts: general
\end{keywords}

%%%%%%%%%%%%%%%%%%%%%%%%%%%%%%%%%%%%%%%%%%%%%%%%%%

%%%%%%%%%%%%%%%%% BODY OF PAPER %%%%%%%%%%%%%%%%%%

\section{Introduction}

The recent discovery of a short gamma-ray burst (GRB) associated with a gravitational-wave event 
from a binary neutron star merger (e.g., \citealp{abbott2017}) has sparked renewed interest in these fascinating 
phenomena. GRBs have been studied for decades and great progress has been made in understanding them
(see, e.g., \citealp{kumarandzhang} for a recent review). However, a general consensus on the prompt variable 
gamma-ray generation mechanism in GRBs has not been reached. This issue continues to be of utmost 
importance, especially as we enter the era of multi-messenger astronomy with the discovery of 
more gravitational-wave signals. Reaching a complete picture of the particular systems that produce
a GRB in connection with gravitational-wave signals will only be reached once the issue of the GRB prompt emission is settled.

The non-thermal quality of the prompt GRB emission spectrum naturally suggests that the radiation is
produced by synchrotron emission from a power-law distribution of electrons (e.g., \citealp{katz1994, reesandmeszaros1994, sarietal1996}). However, the synchrotron mechanism faces the well-known 
``line of death" problem (e.g., \citealt{Preece1998}) observed in many bursts (e.g., \citealt{Preece2000,Ghirlanda2002,Kaneko2006,Nava2011}).
Namely, in the majority of bursts, the measured low-energy spectral slope is significantly harder than expected for synchrotron in the fast cooling regime (which is the expected cooling regime in prompt GRBs). A fast cooling synchrotron slope would also over-produce optical and X-ray emission as compared with upper limits from observations during the prompt phase \citep{BP2014}.
Furthermore, in almost half of the cases, the observed low-energy spectral slope is even harder than the slow cooling spectral slope, where $F_{\nu}\propto \nu^{1/3}$. Nevertheless, the large energy coverage of the {\it Fermi} and {\it Swift} satellites have allowed for broad time-dependent spectral analysis of GRB prompt emission and a more comprehensive understanding in which the synchrotron mechanism might not be ruled out (e.g., \citealt{Guiriec2015,Burgess2017,Oganesyan2017b,Oganesyan2017a,2017Ravasio}). 

GRB models in which the radiation is produced at or close to the photosphere, are a natural way of producing harder spectra (e.g., \citealt{Goodman1986,Thompson1994,Meszaros2000,Giannios2006,Peer2006,Beloborodov2010,Lazzati2010,Ryde2010,Giannios2012,Pe'er2012,Pe'er2015}).
These models may also include a significant synchrotron component \citep{BG2017} which due to the small radius of the emitting region, could become self-absorbed near the X-ray band and thus be consistent with the upper limits on the prompt optical and X-rays. The main concern however with all of these models has to do with reproducing the observed variability of GRB light-curves For $R\approx R_{\rm ph}\lesssim 10^{13}$cm, the typical dynamical time-scale is expected to be $t_{\rm dyn}=R/2c\Gamma^2\lesssim 0.01$ s, almost two orders of magnitude smaller than the observed variability (e.g., \citealt{Fishman1995,Norris1996,Quilligan2002}). The variability and temporal evolution in these models must then be provided by the central engine activity or the propagation of the jet through the stellar envelope. An additional concern has to do with the ``early steep decay"  radiation observed at the end of the prompt phase in many GRBs (e.g, \citealt{Tagliaferri2005}), where the X-ray luminosity of the burst is seen to decline as a power-law with a decay index between 3 to 5. This decline is naturally accounted for when $t_{\rm dyn}\approx t_{\rm v}$ by a purely geometrical effect, high-latitude emission \citep{KP2000}. Of course, shallower declines are also possible, if the time-scale for the shutting down of the engine and/or the dissipation process are long enough \citep{BarniolDuran2009,Fan2005}\footnote{In fact, declines that are more rapid than the regular high-latitude emission are also possible if the prompt radiation is anisotropic in the co-moving frame \citep{BarniolDuran2016,BG2016}.}. Instead, for photospheric-like models, the high latitude emission decays too fast ($t_{\rm dyn}\ll t_{\rm GRB}$) and the early steep decline must be produced by the shutting down of the central engine, by the dissipation process, or by some combination of the two. As shown by \cite{BGM2017}, at least in the case of magnetar central engines, where more robust predictions can be made, this does not seem to occur naturally.

In this paper we focus on the synchrotron mechanism as the origin of the variable prompt $\gamma$-ray emission in GRBs.
We revisit it by obtaining several general constraints on any synchrotron GRB model based on typical observed 
properties of the prompt emission of GRBs. We focus on ``marginally fast cooling" conditions \citep{Daigne2011} that can allow for a hard low-energy spectral slope, while maintaining high efficiency. By marginal fast cooling we mean that electrons cool on a timescale similar to the dynamical time at the source. This paper partially follows the work of \cite{pawananderin, pazandtsvi, BP2014} 
and we recover many of their results. However, we make use of results of particle-in-cell (PIC) simulations 
\citep{SironiSpitkovsky2014,Kagan2015,Guo2015,SKL2015,Sironi2015,Werner2016} to guide our efforts. In particular, as these studies show that a significant fraction of particles are expected to be accelerated to large energies in both shocks and reconnection, we can strongly limit the energy per particle at the emitting region, and strongly disfavour non-magnetic jets as the origin of the required emission. Furthermore, as PIC simulations of highly magnetized dissipation regions show that the emitting plasmoids may exhibit relativistic motions compared to the bulk frame of the jet (the jet co-moving frame), we relax the assumption made in previous studies that the emitters' Lorentz factor (LF) equals the bulk LF. Indeed we find that $\Gamma_{\rm em}>\Gamma_b$, i.e. some relativistic motion of the emitters, relative to the bulk frame, is necessary under these conditions.

The paper is organized as follows. In \S \ref{sec:General} we discuss general constraints on the required conditions for synchrotron to account for the prompt emission, in terms of the energy per particle, the cooling regime, implications on the jet composition and LF of the emitting material and the contribution of Inverse Compton. Motivated by our results in this section, we turn in \S \ref{sec:Reconnection} to discuss magnetic reconnection models. In these models, the allowed parameter range is further constrained and, given the large required values of the magnetization upstream of the emitting region, the particle spectra may become harder than $dN/d\gamma\propto \gamma^{-2}$. Nonetheless, self-consistent solutions that satisfy all the observational constraints are still available. In \S \ref{sec:Dis} we explore a variant of the basic model, and also provide a general discussion of the spectral shape and compare our results with previous studies. We present our conclusions in \S \ref{sec:conclusion}.

\section{General constraints}
\label{sec:General}
Let us consider some general constraints on the synchrotron emission mechanism in the context of the prompt emission of GRBs. We assume that the energy available per electron is $E$, and that the thermal (or random) LF of these electrons is given by 
\beq
\gamma_{\rm e} \equiv \frac{E}{m_{\rm e} c^2} = 2 \times 10^3 E_{\rm GeV},
\label{gamma_e}
\eeq
where $E_{\rm GeV}$ is the energy in units GeV, $m_{\rm e}$ is the electron mass and $c$ is the speed of light.

We initially focus on two extreme situations. First, we consider the instantaneous injection of this energy to the electrons.  Second, we consider a simple model in which electrons receive this energy continuously over a dynamical time: the ``slow heating" model. We note that when referring to the thermal (or random) LF of the electrons relative to the emitting region we will use $\gamma$, whereas the emitting region is assumed to be moving at a LF $\Gamma_{\rm em}$ towards the observer. 

\subsection{Instantaneous injection}
\label{sec:inst}
We assume here that electrons are accelerated instantaneously by some mechanism (e.g., ``one-shot" shock acceleration). Both relativistic shocks and magnetic reconnection can accelerate particles ``instantaneously" (i.e. over a time-scale much shorter than the dynamical time) and over a broad range of LFs. In shocks, the power-law  index of the particles' LF distribution is typically expected to be $p>2$ (e.g., \citealt{Heavens1988,Bednarz1998,Achterberg2001}). This leads to a distribution in which the minimal value of $\gamma$ dominates both the total energy stored in the electrons (which scales as $E_{\rm tot}\propto\gamma^{2-p}$) and their overall number (which scales as $N_{\rm tot}\propto\gamma^{1-p}$). The same holds true also for magnetic reconnection models with $\sigma \lesssim 10$ (e.g.,  \citealt{Cerutti2012,SironiSpitkovsky2014,Guo2014,Melzani2014}). Here we explore instantaneous acceleration with $p>2$. We turn to more gradual heating models in \S \ref{sec:slowheat} and to models with $p<2$ in \S \ref{sec:particledist}.

Under these assumptions, each electron gains an energy $E$ and therefore its instantaneously attained LF is 
\beq
\gamma_{\rm i} = \gamma_{\rm e},
\label{gamma_i}
\eeq
given by equation (\ref{gamma_e}). For general acceleration mechanisms (i.e., not necessarily instantaneous), this will serve as an upper limit on the LF that electrons can achieve for a given energy per particle, $E$. Electrons with LF $\gamma_{\rm i}$ in a co-moving magnetic field strength $B_{\rm em}$ radiate via the synchrotron process at a characteristic energy given by 
\beq
\nu_{\rm p} = \frac{e B_{\rm em} \gamma_{\rm i}^2 \Gamma_{\rm em}}{2 \pi m_{\rm e} c},
\eeq
Observationally, the peak of the GRB in the source frame is approximately $\nu_{\rm p} \sim 300$ keV (e.g., \citealt{Preece2000,Kaneko2006,Nava2011}). Furthermore compactness arguments show that $\Gamma_{\rm em}\geq 100$ (e.g, \citealt{Fenimore1993,Woods1995,Lithwick2001}). With these constraints, the magnetic field is 
\beq
B_{\rm em} \approx \frac{2.6 \times 10^{11} \nu_{\rm p,5.5}}{\gamma_{\rm i}^2 \Gamma_{\rm em,2}} \approx \frac{(6 \times 10^4 \, {\rm G}) \, \nu_{\rm p,5.5}}{\Gamma_{\rm em,2} E_{\rm GeV}^2},
\label{B_field}
\eeq
where we made use of equation (\ref{gamma_i}) and $\nu_{\rm p,5.5}$ is the (source frame) peak frequency in units of 300 keV. We have adopted here the usual convention ($Q_n = Q / (10^n {\rm cgs})$).

The (source frame) variability time of a single pulse in the GRB prompt emission light curve is of the order of $t_{\rm v} \sim 0.5$ s (e.g., \citealt{Fishman1995,Norris1996,Quilligan2002}). The cooling LF is defined as the LF for which the synchrotron cooling time is equal to the dynamical time. Here, the dynamical time in the co-moving frame is $t_{\rm v}'\sim t_{\rm v} \Gamma_{\rm em}$ and thus
\bea
\gamma_{\rm c} &=& \frac{6 \pi m_{\rm e} c}{\sigma_{\rm T} B_{\rm em}^2 t_{\rm v} \Gamma_{\rm em}}=4 \times 10^{-3}\, \frac{\Gamma_{\rm em,2} E_{\rm GeV}^4}{t_{\rm v,0.5} \, \nu_{\rm p,5.5}^2 }.
\label{cooling_LF}
\eea
Comparing $\gamma_{\rm c}$ with $\gamma_{\rm i}$ we find 
\beq
\frac{\gamma_{\rm c}}{\gamma_{\rm i}} \approx 2\times 10^{-6}\frac{\Gamma_{\rm em,2} E_{\rm GeV}^3}{t_{\rm v,0.5} \, \nu_{\rm p,5.5}^2 }.
\eeq

Synchrotron models for prompt GRBs require that $\gamma_{\rm c}/\gamma_{\rm i}\approx 1$. This is because $\gamma_{\rm i}\gtrsim \gamma_{\rm c}$ is needed in order to account for the large observed efficiency of GRBs (e.g., \citealt{Fan2006,Beniamini2015,Beniamini2016}), while $\gamma_{\rm c} \gtrsim \gamma_{\rm i}$ is needed in order to avoid a low-energy spectral slope that is too soft as compared with observations, i.e the ``line of death" problem (e.g., \citealt{Preece2000,Ghirlanda2002,Kaneko2006,Nava2011}), and an excess of X-ray and optical emission as compared with observational limits \citep{BP2014}. While this can be achieved for some choice of $\Gamma_{\rm em}, E_{\rm GeV}$ (see purple region in figure \ref{figgcgm} and also \citealt{Daigne2011,pazandtsvi}), it is not clear why this would be the case in most GRBs. This issue could be resolved in a slow heating scenario, since particles maintain their energy (and thus emitting frequency) over a long time and therefore the fast cooling spectrum and excess low-energy emission can be avoided.
Various studies have considered such continuous acceleration models in the past (e.g.,  \citealt{Ghisellini1999,gianniosandspruit2005,pawananderin,asano2009,Fan2010,Daigne2011,BP2014}; see also discussion in \S \ref{sec:compare}). In what follows, we reconsider this possibility in a slightly different context.

\subsection{Slow heating}
\label{sec:slowheat}
\begin{figure}
\begin{center}
\includegraphics[scale=0.5]{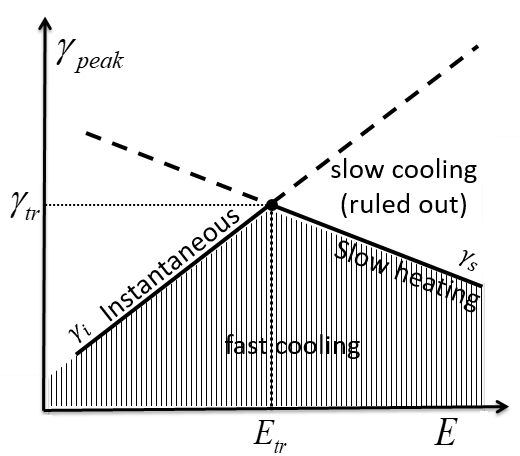}
\end{center}
\caption{The typical LF of the electrons radiating at the observed peak synchrotron energy, $\gamma_{\rm peak}$, versus the available energy per electron, $E$. Both the instantaneous and the slow heating cases are shown. All solutions must satisfy $\gamma_{\rm peak}\leq\gamma_i$ or else they would require a LF larger than allowed by the available energy per electron. In particular, slow heating solutions are not available below $E_{\rm tr}$. Similarly, no self-consistent solutions are available for $\gamma_{\rm peak}>\gamma_{\rm s}$ (see \S \ref{sec:intermediate}). Below the slow heating line, electrons are fast cooling due to synchrotron radiation. Viable synchrotron models for the GRB emission lie close to the solid ``slow heating" line, where electrons are only marginally fast cooling. }
\label{fig1}
\end{figure}

In the opposite extreme case, instead of being accelerated instantaneously, particles could be heated ``slowly" over a dynamical time. In this case, the heating rate is
\beq
\dot{\epsilon_{\rm h}} = \frac{E}{\Gamma_{\rm em} t_{\rm v}} \approx 3\times 10^{-5} \frac{E_{\rm GeV}}{\Gamma_{\rm em,2} \, t_{\rm v,0.5}} \mbox {erg s}^{-1}.
\label{eq:heatrate}
\eeq
The last expression assumes that all the particles are heated throughout the entirety of the dynamical time. In \S \ref{sec:shortheat} we explore the possibility of acceleration over a shorter timescale, in \S \ref{sec:intermediate} we explore the possibility of multiple acceleration episodes, and in \S \ref{sec:particledist}, we explore models with $p<2$, in which effectively only a selective fraction of particles are heated. 
As they are heated, particles also cool via the synchrotron process with a cooling rate that is given by
\beq
\dot{\epsilon_{\rm c}} = \frac{1}{6 \pi} \sigma_{\rm T} c \gamma_{\rm s}^2 B_{\rm em}^2 \approx 10^{-15} \gamma_{\rm s}^2 B_{\rm em}^2
\approx \frac{6.7 \times 10^{7} \, \mbox {erg s}^{-1}}{\Gamma_{\rm em,2}^2 \gamma_{\rm s}^2},
\label{eq:coolrate}
\eeq
where $\gamma_{\rm s}$ is their electron LF. The last expression was obtained using equation (\ref{B_field}), therefore, it incorporates the constraint on the observed synchrotron peak energy. We now set the cooling rate equal to the heating rate $\dot{\epsilon_{\rm h}} = \dot{\epsilon_{\rm c}}$ and find 
\beq
\gamma_{\rm s} \approx 1.5 \times 10^6 \frac{ t_{\rm v,0.5}^{1/2} \, \nu_{\rm p,5.5}}{\Gamma_{\rm em,2}^{1/2}  E_{\rm GeV}^{1/2}}.
\label{gamma_s}
\eeq
Given that $\gamma_{\rm e}$ is the LF that corresponds to the energy available per electron (and also the LF attained in the instantaneous injection case, $\gamma_{\rm i}$), the {\it true} electron LF in the slow heating case cannot be larger than $\gamma_{\rm i}$ (see \S \ref{sec:inst}).
Since $\gamma_e\propto E$ while $\gamma_s\propto E^{-1/2}$, there is a minimum `transition' energy, $E_{\rm tr}$, for which $\gamma_s\leq \gamma_e$ and slow heating solutions become available. $E_{\rm tr}$ and its corresponding (maximal) LF, $\gamma_{\rm tr}$, are 
\bea
E_{\rm tr} &\approx& 83 \, {\rm GeV}\frac{t_{\rm v,0.5}^{1/3} \, \nu_{\rm p,5.5}^{2/3} }{\Gamma_{\rm em,2}^{1/3} }
\label{eq:Etr}
\eea
\bea
 \gamma_{\rm tr} &\approx& 1.7 \times 10^5 \frac{t_{\rm v,0.5}^{1/3} \, \nu_{\rm p,5.5}^{2/3}}{\Gamma_{\rm em,2}^{1/3}}\equiv\gamma_{\rm i}^{1/3}\gamma_{\rm s}^{2/3}, \label{eq:gtr}
\eea
which weakly depend on the observables and $\Gamma_{\rm em}$ (in \S \ref{sec:revisedlimits} we provide slightly revised versions of these equations, taking into account our constraints on $\Gamma_{\rm em}$, derived in \S \ref{sec:luminosity}). In figure \ref{fig1} we present the electrons LF in the instantaneous injection case (equation (\ref{gamma_e})) and in the slow heating case (equation (\ref{gamma_s})).

When $E=E_{\rm tr}$ the cooling LF, defined in equation (\ref{cooling_LF}), is $\gamma_{\rm c} = \gamma_{\rm tr}$ (see figure \ref{figgcgm}). The reason is that at $E_{\rm tr}$, by definition, $\gamma_{\rm s} = \gamma_{\rm i}$, and since equation (\ref{gamma_s}) both (i) incorporates the constraint on the observed peak synchrotron energy (equation (\ref{B_field})), and (ii) assumes that heating occurs in a dynamical time, then it follows that $\gamma_{\rm c} = \gamma_{\rm i}$ and thus $\gamma_{\rm c} = \gamma_{\rm s} = \gamma_{\rm tr}$, when $E=E_{\rm tr}$.

%%%%%%%%%%%%%%%%Figure%%%%%%%%%%%%%%%%%
%%%%%%%%%%%%%%%%%%%%%%%%%%%%%%%%%%%

\subsection{Intermediate regime}
\label{sec:intermediate}
\begin{figure}
\begin{center}
\includegraphics[scale=0.39]{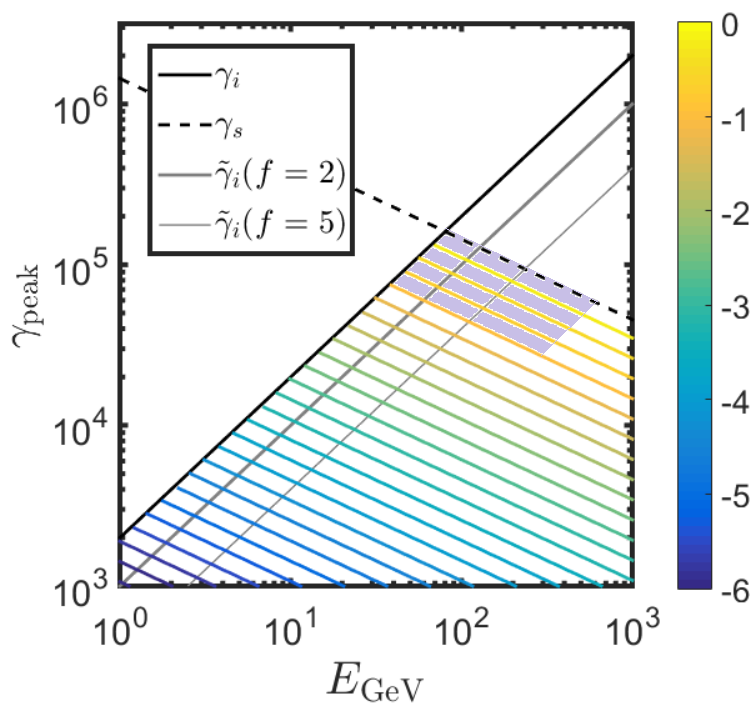}
\end{center}
\caption{ $\log_{10}(\gamma_{\rm c}/\gamma)$ as implied by the association $\nu_{\rm p}=\nu(\mbox{max}(\gamma,\gamma_{\rm c}))$. Self consistent solutions are only available when $\nu_{\rm p}=\nu(\gamma)$, i.e. when electrons are fast cooling. Grey lines depict the injected LF of typical electrons, given that they are accelerated $f$ times within $t_v$. Marginally fast cooling solutions require $\gamma\approx\gamma_{\rm s}$. Below this line, the spectrum is in strong contradiction with observations. The purple shaded region depicts the allowed parameter space for marginally fast cooling synchrotron solutions (taken here as $0.1<\gamma_c/\tilde{\gamma}_i<1$). The upper limit on $E_{\rm GeV}$ arises from the requirement that IC is sub-dominant as compared with synchrotron (see \S \ref{sec:IC}).}
	\label{figgcgm}
\end{figure}

Consider now the intermediate regime between instantaneous acceleration and continuous heating. In this regime the particles can be thought of as having undergone $f>1$ acceleration episodes, with each acceleration boosting them to a LF 
\begin{equation}
\tilde{\gamma}_{\rm i}=\gamma_{\rm i}/f. 
\end{equation} This allows for solutions characterized by a LF smaller than $\gamma_{\rm e}$, but without necessarily invoking continuous acceleration (see grey lines in figure \ref{figgcgm}).
The shorter time between consecutive acceleration episodes effectively reduces the available cooling time by a factor $f$ and thus increases $\gamma_{\rm c}$ by the same factor \citep{pawananderin,BP2014} as compared with equation (\ref{cooling_LF}). Below, we examine the allowed parameter space and implications corresponding to solutions with $f\geq 1$.

Consider first the case in which electrons are fast cooling. In this case the peak of the synchrotron spectrum is produced by electrons with $\tilde{\gamma}_{\rm i}$. Since $\gamma_{\rm c}\propto B_{\rm em}(\gamma_{\rm peak})^{-2}\propto \gamma_{\rm peak}^{4}$ and since when $\gamma_{\rm peak}=\gamma_{\rm s}$ we have $\gamma_{\rm c}=\gamma_{\rm s}$ (see \S \ref{sec:slowheat}), we can re-write equation (\ref{cooling_LF}) as a scaling relation $\gamma_{\rm c}=(\gamma_{\rm peak}/\gamma_{\rm s})^4 \gamma_{\rm s}$. It follows that
\beq
\frac{\gamma_{\rm c}}{\gamma_{\rm peak}}=\bigg(\frac{\gamma_{\rm peak}}{\gamma_{\rm s}}\bigg)^3\ll1.
\label{eq:gcgi}
\eeq
Therefore, solutions with $\gamma_{\rm peak}=\tilde{\gamma}_{\rm i}<\gamma_{\rm s}$ are strongly fast cooling, even when the acceleration is done in multiple episodes.

For $\tilde{\gamma}_i=\gamma_{\rm s}$, electrons radiating at the peak become slow cooling, i.e. $\tilde{\gamma}_i=\gamma_{\rm peak}=\gamma_{\rm s}=\gamma_{\rm c}$. At these conditions
equation (\ref{gamma_e}) no longer directly constrains the minimum injected LF, $\gamma_{\rm m}$, as electrons with $\gamma_{\rm c}$ carry more energy overall than those with $\gamma_{\rm m}$ (assuming the slope of the electrons' injected number spectrum, $dN/d\gamma\propto \gamma^{-p}$, is $p<3$). Therefore solutions with $\tilde{\gamma}_i=\gamma_{\rm s}$ can be either marginally fast or slow cooling. Nonetheless, strongly slow cooling solutions are disfavoured as the radiative efficiency of the GRB is reduced significantly at those conditions (the efficiency scales as $(\gamma_{\rm c}/\gamma_{\rm m})^{2-p}<1$), while GRB observations show that the radiative efficiency of GRBs must be large. Notice that solutions with $\tilde{\gamma}_i=\gamma_{\rm peak}>\gamma_{\rm s}$ are impossible. This is because: (a) No fast cooling solutions exist in this regime (if $\gamma_{\rm peak}>\gamma_{\rm s}$ then it is not possible to have $\gamma_{\rm c}<\gamma_{\rm peak}$, see equation (\ref{eq:gcgi})), (b) For slow cooling:
\beq
\tilde{\gamma}_i=\bigg( \frac{\gamma_{\rm c}}{\gamma_{\rm m}}\bigg)^{1-p}\gamma_{\rm c}<\gamma_{\rm c},
\eeq
while as shown above, slow cooling requires $\gamma_{\rm c}=\gamma_{\rm s}$. Combining these two relations we have $\tilde{\gamma}_i<\gamma_{\rm s}$ contrary to our initial assumption. Thus, no solutions exist with $\gamma_{\rm peak}>\gamma_{\rm s}$. In particular, this implies that $\gamma_{\rm tr}$ is the largest allowed value of the electrons' LF (see figure \ref{fig1}) and correspondingly $B_{\rm em}(\gamma_{\rm tr}) $ is the lowest allowed value of the magnetic field.

Figure \ref{figgcgm} depicts the value of $\gamma_{\rm c}/\gamma$ in the $\gamma-E$ plane. Allowed solutions correspond to $\gamma_{\rm peak}\leq\gamma_{\rm s}$, for which electrons are fast (or at best, marginally fast) cooling. We define here the marginally fast cooling regime as $0.1<\gamma_c/\tilde{\gamma}_i<1$ (shown as a purple region in figure \ref{figgcgm}). This is in accord with recent observational results suggesting a synchrotron cooling break with $\nu_{\rm p}/\nu_{\rm c}\approx 5-80$ \citep{Oganesyan2017b}. Since for $\tilde{\gamma}_i\gg\gamma_c$ the GRB spectrum is in strong contradiction with observations ('fast cooling line of death'), we focus on the {\it marginally fast cooling solutions} \citep{Daigne2011} with  $\tilde{\gamma}_{\rm i}\approx\gamma_{\rm c}$, which implies $\gamma_{\rm peak}\approx\gamma_{\rm s}$.

%%%%%%%%%%%%%%%%Figure%%%%%%%%%%%%%%%%%
%%%%%%%%%%%%%%%%Figure%%%%%%%%%%%%%%%%%

\subsection{Implications for the jet's composition}
\label{sec:MatterorMag}
The initial energy per baryon in the jet, $\eta$, can be related to $E\Gamma_{\rm em}/(m_p c^2)$, leading to $\eta \approx 110 \Gamma_{\rm em,2} E_{\rm{ GeV}}$, and demonstrating that large initial amounts of energy per baryon are required (given that $E>E_{\rm tr}$); see equation (\ref{eq:Etr}). Notice that the lower limits become even larger, by a factor of $m_p/m_e$, for a pair dominated outflow. In \S \ref{sec:luminosity}, we show that the observed luminosity implies that the LF of the emitting material must be very large, $\Gamma_{\rm em} \approx 1.8\times 10^4E_{\rm{ GeV}}^{-1/3}$, resulting in even larger energies per baryon at the base of the jet $\eta \approx 2\times 10^4 E_{\rm{ GeV}}^{2/3}$.

At the emission region, the energy per particle in the co-moving frame depends on the nature of the jet (baryonic or Poynting flux). For a baryonic jet in which dissipation proceeds through internal shocks we expect $E=\epsilon_e g m_pc^2\lesssim$ GeV (where $\epsilon_e\leq 1$ is the fraction of the total energy that goes to accelerating electrons and $g\sim 2-3$ is a numerical factor that depends on the relative LF of the colliding shells). However, as shown above, this leads to strongly fast cooling conditions with $\gamma_{\rm c}/\gamma_{\rm i}\lesssim 10^{-5}$. This in turn translates to a spectral slope that is in stark contradiction with observations. The large required values of $E$, suggest that the jet must be magnetically dominated. Assuming that the magnetic energy can be efficiently dissipated at the emitting region, $E$ is related to the magnetization of the jet in the upstream of the radiation zone, $\sigma_{\rm up}=B_{\rm b}^2/4\rho' c^2$ (where $B_{\rm b}$ is the magnetic field, and $\rho'$ is the matter density, both measured in the bulk frame):
	\beq
	E\leq \epsilon_e \sigma_{\rm up} m_p c^2=0.17 \epsilon_{e,0.2}\sigma_{\rm up} \rm{ GeV}
	\label{Esigma}
	\eeq
	where $\epsilon_{e,0.2}=\epsilon_e/0.2$. Assuming $\epsilon_e\approx 0.2$, this means that $\sigma_{\rm up}\gtrsim 100$ is required in order to have $E\gtrsim 20$ GeV, as needed to account for marginally fast cooling solutions (see equation (\ref{eq:Etr}) and also equation (\ref{eq:Etrplas}) below).

Therefore, marginally fast cooling synchrotron implies jets that have both a very large amount of energy per baryon at their base, {\it and} just before the emission zone.

\subsection{Observed Luminosity - Determining the Lorentz factor of the emitting material}
\label{sec:luminosity}
The observed isotropic equivalent luminosity of GRBs, $L_{\rm rad}\approx 10^{52} \mbox{erg s}^{-1}$ introduces a constraint on the Lorentz factor of the emitting material, $\Gamma_{\rm em}$. We focus here on emitting regions that are quasi spherical in the co-moving frame. In appendix \ref{sec:planar} we show that the conclusion regarding the LF of the emitting material holds also for planar geometry, as applicable for instance to internal shock models.

Consider a spherical emitting region in the co-moving frame. In its own frame, the emitting region grows with time at a speed $v_g<c$. The region grows over the dynamical time-scale, which is observationally related to the variability time-scale $t_{\rm v}'$. In marginally fast cooling conditions, the radiated energy, is proportional to the energy of electrons in this volume. As usual, we denote by $\epsilon_e,\epsilon_B$ the fractions of the dissipated energy deposited in electrons and magnetic fields respectively. In the co-moving frame, the emission lasts over a period $t_{\rm v}'=\Gamma_{\rm em}t_{\rm v}$. The observed luminosity is then 
\beq
L_{\rm obs}=\Gamma_{\rm em}^4 L'=\Gamma_{\rm em}^4 e_e'\frac{4\pi t_{\rm v}'^3v_g^3}{3t_{\rm v}'} =\Gamma_{\rm em}^6 \frac{\epsilon_e}{\epsilon_B}\frac{4\pi}{3}\frac{B_{\rm em}^2}{4\pi} t_{\rm v}^2v_g^3,
\eeq
where $e_e'\equiv \frac{\epsilon_e}{\epsilon_B}\frac{B_{\rm em}^2}{4\pi}$ is the co-moving energy density in electrons.
Using the magnetic field implied by the condition of marginally fast cooling $B_{\rm em}=B_{\rm em}(\gamma=\gamma_{\rm s})$, we get an estimate of the Lorentz Factor:
\beq
\Gamma_{\rm em}=1.5\times 10^4 \bigg(\frac{\epsilon_B}{\epsilon_e}\bigg)^{1/6}v_{g,.3c}^{-1/2}\frac{\nu_{\rm p,5.5}^{1/3} L_{52}^{1/6}}{E_{\rm GeV}^{1/3}},
\label{eq:Gammaspher}
\eeq
where $v_{g,.3c}\equiv v_g/0.3c$. Interestingly the dependence on the observed parameters, and on $\epsilon_B/\epsilon_e$ is very weak. Efficiency consideration imply $10^{-2}<\epsilon_B/\epsilon_e<10$ ($\epsilon_e$ has to be large enough to tap a significant of energy from the electrons, while $\epsilon_B$ cannot be too small in order to avoid most of this energy being deposited in an unobserved SSC component, see equation (\ref{eq:upperE}) in \S \ref{sec:IC} and \citealp{pazandtsvi}). We find that even for weakly magnetized jets, unless $E_{\rm GeV}>10^3$, the LF of the emitting material is $\Gamma_{\rm em}\gtrsim 700$, which is incompatible with estimates of the jet's bulk LF of typical bursts, as obtained for instance by early afterglow peaks (e.g., \citealp{Liang2010,Lu2012,Ghirlanda2012}). We note however that this derivation does not explicitly assume that $\Gamma_{\rm em}$ is the bulk Lorentz factor, $\Gamma_b$. In fact, this consideration shows that $\Gamma_{\rm em}\gg \Gamma_b$. Thus, {\it marginally fast cooling requires relativistic motions in the bulk frame.}

A lower limit on the emission radius arises from requiring that the emitting blob has sufficient time to grow, i.e., from requiring that the co-moving expansion time $t_{\rm exp}'= R/c\Gamma_b$ of the shell is larger than $t_{\rm v}'$, the lifetime of the emitting region. This leads to
	\bea
	& R\geq 2 \Gamma_{\rm em} \Gamma_b c t_{\rm v} \nonumber \\ &=1.8 \times 10^{17} \frac{\Gamma_{b,2.5} \nu_{\rm p,5.5}^{1/3} L_{52}^{1/6}t_{\rm v,0.5}}{E_{\rm GeV}^{1/3}v_{g,.3c}^{1/2}}\bigg(\frac{\epsilon_B}{\epsilon_e}\bigg)^{1/6} \mbox{cm.}
	\label{eq:rspher}
	\eea
Using the value of $E_{\rm GeV}$ inferred by the requirement for marginally fast cooling solutions, the emitting radius in equation (\ref{eq:rspher}) is quite large but still marginally consistent with upper limits implied by the deceleration radius ($R\lesssim 10^{17}\mbox{cm}$).

In equation (\ref{eq:rspher}) we have assumed that $\Gamma_{\rm em}\gg \Gamma_b$, which effectively suggests relativistic motion in the bulk frame. If this is not the case, and instead $\Gamma_{\rm em}=\Gamma_b$, then the lower limit on the LF of the emitting material as given by equation (\ref{eq:Gammaspher}) would lead to a very large lower limit on the radius:
\beq
R\geq 2 \Gamma_{\rm em}^2 ct_{\rm v} =1.2\times 10^{19} \bigg(\frac{\epsilon_B}{\epsilon_e}\bigg)^{1/3}\frac{\nu_{\rm p,5.5}^{2/3}L_{52}^{1/3}}{E_{\rm GeV}^{2/3}v_{g,.3c}}t_{\rm v,0.5} \mbox{ cm}.
\eeq
As mentioned above, these values are inconsistent with upper limits implied by the deceleration radius (even when redshift corrections are applied). This consideration, together with the estimates of $\Gamma_{\rm em}$ above lead us to conclude that $\Gamma_{\rm em}\gg \Gamma_b$ is required, and therefore that relativistic motion in the bulk frame is inevitable unless the energy per particle is unrealistically large: $E>10^3$ GeV. We show below (\S \ref{sec:IC}) that values of $E\gtrsim700$GeV (in models involving an acceleration of a significant fraction of the electrons) are in fact inconsistent with the synchrotron models considered in this paper, as they would lead to excess IC emission and cooling. Furthermore, as will be shown in \S \ref{sec:relmotion}, in magnetic reconnection models there are two additional independent considerations that both reach the same outcome of relativistic motion in the bulk frame.

\subsection{Optical depth of the emitting region and Synchrotron self-Compton}
\label{sec:IC}
The conditions required to account for a balanced heating synchrotron model for GRBs' prompt emission discussed in \S \ref{sec:General} impose constraints on $\tau$, the optical depth of the emitting region to Thomson scatterings. $\tau$ can be related to the co-moving electron number density, $n'$, of the emitting region via \citep{Abramowicz1991}
\beq
\tau=\int \Gamma_{\rm em}(1-\beta_{\rm em})\sigma_T n' dr \approx  \sigma_T n' \frac{v_g t_v \Gamma_{\rm em}}{2},
\eeq
where the integral is over the length of the emitting region and we have used $\Gamma_{\rm em} \gg 1$. Using the constraint for $\Gamma_{\rm em}$ in equation (\ref{eq:Gammaspher}), we use equation (\ref{gamma_s}) to obtain $\gamma_{\rm s}$ (equation (\ref{eq:gammasplasmoid}) below) and $B_{\rm em}=B(\gamma_{\rm s})$. $n'$ can be related to the energy density in electrons and the energy of an individual electron at any given time:
\beq
n'=\frac{\epsilon_e}{\epsilon_B}\frac{B_{\rm em}^2}{4\pi}\frac{1}{\gamma_{\rm s} m_e c^2}=2 \times 10^{-3} \frac{\epsilon_{e,0.2}^{11/12}E_{\rm GeV}^{7/3}L_{52}^{1/12}}{\epsilon_B^{11/12}\nu_{\rm p,5.5}^{17/6}t_{\rm v,0.5}^{2.5}v_{g,.3c}^{1/4}}\mbox{ cm}^{-3}.
\eeq
With this, the corresponding optical depth is then
\beq
\tau=6\times 10^{-14}\frac{\epsilon_{e,0.2}^{3/4}E_{\rm GeV}^{2}L_{52}^{1/4}v_{g,.3c}^{1/4}}{\epsilon_B^{3/4}\nu_{\rm p,5.5}^{5/2}t_{\rm v,0.5}^{1.5}} \ll 1.
\eeq
Clearly $\tau \ll 1$ for any reasonable value of $E_{\rm GeV}$. This leads to important conclusion,  that if there is also a photospheric component present in GRBs' prompt emission, it cannot originate from the same location as the synchrotron emission in the balanced-heating models considered in this paper. 

The optical depth can also be related to the Compton-$Y$ parameter, which in the Thomson regime is given by
\beq
\label{eq:Y}
Y_{\rm Th}\approx \tau \gamma_{\rm s}^2= 10^{-3} \frac{\epsilon_{e,0.2}^{11/12}E_{\rm GeV}^{4/3}L_{52}^{1/12}v_{g,.3c}^{3/4}}{\epsilon_B^{11/12}\nu_{\rm p,5.5}^{5/6}t_{\rm v,0.5}^{1/2}}\ll 1.
\eeq
Since typically $\gamma_{\rm s}h\nu_p>\Gamma_{\rm em}m_e c^2$, $Y_{\rm Th}$ may be further suppressed due to the Klein Nishina effect. Assuming marginally fast cooling, where the slope of $F_{\nu}$ below the peak is $\nu^{1/3}$ (see discussion in \S \ref{sec:specshape}), this suppression factor can be approximated by \citep{Ando2008}:
\bea
\label{zetaKN}
&\zeta_{\rm KN} \approx \min\bigg[\bigg(\frac{\Gamma_{\rm em} m_e c^2}{\gamma_{\rm s} h \nu_p}\bigg)^{\frac{4}{3}},1\bigg]\\ \nonumber &=\min\bigg[0.16\frac{L_{52}^{1/3}\epsilon_B^{1/3}}{\epsilon_{e,0.2}^{1/3}\nu_{\rm p,5.5}^{2}t_{\rm v,0.5}^{2/3}v_{g,.3c}},1\bigg]
\eea
The final value of the Compton parameter is then $Y=Y_{\rm Th}\zeta_{\rm KN}$. The power of the synchrotron self-Compton and the cooling time of electrons due to IC, are both reduced by a factor $Y$ as compared with the synchrotron power and cooling time. Equations (\ref{eq:Y}) and (\ref{zetaKN}) imply that in order to keep the IC suppression small ($Y\lesssim 1$), and maintain high efficiency of the sub-MeV synchrotron component as required by observations (e.g, \citealt{Fan2006,Beniamini2015,Beniamini2016}), we must put an upper limit on the allowed value of\footnote{We assume in the following derivation that $\zeta_{KN}\leq 1$ applies. The dependence on parameters changes slightly if this is not the case, but the typical energies are similar.} $E$:
\beq
\label{eq:upperE}
E\lesssim 700 \frac{\epsilon_B^{7/16}\nu_{\rm p,5.5}^{17/8}t_{\rm v,0.5}^{5/4}v_{g,.3c}^{3/16}}{\epsilon_{e,0.2}^{7/16}L_{52}^{5/16}}\mbox{GeV.}
\eeq
When this condition is satisfied, the assumptions that synchrotron dominates the energy release rate and the observed emission are indeed self-consistent.

\subsection{Resulting constraints on the energy per particle and particles' LF}
\label{sec:revisedlimits}
Combining the value of $\gamma_{\rm s}$ as implied by equation (\ref{gamma_s}) with the estimate of $\Gamma_{\rm em}$ given by equation (\ref{eq:Gammaspher}), we find
\beq
\label{eq:gammasplasmoid}
\gamma_{\rm s}=1.1\times 10^5 t_{\rm v,0.5}^{1/2}\nu_{\rm p,5.5}^{5/6} L_{52}^{-1/12}E_{\rm GeV}^{-1/3}\epsilon_{e,0.2}^{1/12}\epsilon_B^{-1/12}v_{g,.3c}^{1/4}.
\eeq
Plugging this back into equations (\ref{eq:Etr}) and (\ref{eq:gtr}) we can rewrite the limits on the minimal allowed energy per particle, the corresponding electrons' (maximal) LF and the (minimum) allowed value for the magnetic field:
\bea
E_{\rm tr}\approx 20 t_{\rm v,0.5}^{3/8}\nu_{\rm p,5.5}^{5/8}L_{52}^{-1/16}\epsilon_{e,0.2}^{1/16}\epsilon_B^{-1/16}v_{g,.3c}^{3/16}  \mbox{GeV}
\label{eq:Etrplas}
\eea
\bea
\gamma_{\rm tr}\approx 4\times 10^4 t_{\rm v,0.5}^{3/8}\nu_{\rm p,5.5}^{5/8}L_{52}^{-1/16}\epsilon_{e,0.2}^{1/16}\epsilon_B^{-1/16}v_{g,.3c}^{3/16}
\label{eq:gtrplas}
\eea
\bea
B_{\rm em}\approx 2 t_{\rm v,0.5}^{-5/8}\nu_{\rm p,5.5}^{-3/8}L_{52}^{-1/16}\epsilon_{e,0.2}^{1/16}\epsilon_B^{-1/16}v_{g,.3c}^{3/16}\mbox{G}.
\label{eq:Btrplas}
\eea
Given the upper limit on the energy per particle implied by equation (\ref{eq:upperE}), we find that the energy per particle (and therefore also the typical LF and the magnetic field) has a range of roughly one and a half orders of magnitude, $20\lesssim E_{\rm GeV} \lesssim 700$, in which it can account for balanced heating synchrotron solutions, as discussed in this paper (the allowed parameter ranges are increased somewhat if we allow for $0.1<\gamma_{\rm c}/\gamma_{\rm s}<1$ as discussed in \S \ref{sec:intermediate}).
\section{Magnetic reconnection models}
\label{sec:Reconnection}
Motivated by the large energy per baryon required by continuous heating or marginally fast cooling models (see \S \ref{sec:MatterorMag}), we consider here specific constraints for magnetically dominated jets.
We focus here on emission from plasmoids, quasi-spherical regions of plasma that have strong magnetic fields and highly energetic particles. These are expected to be the main sources of emission in reconnection models. Such a model has been applied to account for the observed fast flares from blazar jets from active galactic nuclei (e.g., \citealt{Giannios2013,mariaetal2016}).  Previous studies have suggested that these plasmoids could be moving relativistically compared to the bulk frame, as indeed implied by \S \ref{sec:luminosity}.

\cite{mariaetal2016} assume that the particles 
are accelerated instantaneously once they are injected in the plasmoid.
Here, we also consider the possibility that the particles can undergo a slower 
injection of energy, or ``heating", while they reside in the plasmoid during 
major merger events. We assume here that a pulse in the GRB
light-curve arises from the merger of two plasmoids. Naturally,
the most luminous pulses will correspond to the merger of some of the 
largest plasmoids. When two large plasmoids merge into one, the whole
structure relaxes to a new MHD equilibrium. This equilibrium
is reached after an Alfven crossing time, which up to a numerical
factor of order $\lesssim 3$ is of the order of the light-crossing time.
This is also the time during which the energy is released. Such a
merger could excite Alfvenic turbulence, which dissipates energy and heats up
the electrons until they reach a LF, $\gamma_{\rm s}$, where the synchrotron
cooling is balanced by heating (see, e.g., \citealt{Thompson1994}).

Magnetic reconnection simulations find that plasmoids grow in size with velocity $v_g\approx 0.3c$ (e.g., \citealt{Guo2015,Liu2015}). Furthermore, the conditions in the plasmoid can be approximated by $\epsilon_B=1, \epsilon_e=0.2$ \citep{Sironi2015}.
As will be shown below, these values constrain the degree of relativistic motion in the bulk frame as well as the 
power-law distribution of accelerated electrons.
\subsection{Relativistic motion in the bulk frame}
\label{sec:relmotion}
We have seen in \S \ref{sec:luminosity} that the observed luminosity constrains the LF of the emitting material, $\Gamma_{\rm em}$. The large required values of $\Gamma_{\rm em}$, suggest relativistic motion of the emitting material in the bulk frame, with 
\beq
\label{eq:Gammapdef}
\Gamma' \approx \Gamma_{\rm em}/\Gamma_b.
\eeq
Here we show that $\Gamma'$ is strongly restricted by the available Poynting luminosity of the jet, $L_B$.

The (isotropic equivalent) Poynting luminosity of the jet at a radius $R$ is given by:
\beq
\label{eq:LB}
L_B=4\pi R^2\Gamma_b^2\frac{B_{\rm b}^2}{4\pi}c
\eeq
where $B_b$ is the magnetic field in the bulk frame. Since this is the source of energy that feeds the emitters, we must have that $L_B \gtrsim \langle L_{\rm rad}\rangle$. Furthermore, since the `filling factor' of GRB light-curves is of order unity (i.e. there are no prolonged episodes where the luminosity dips below the values typically seen during the $\gamma$-ray pulses), we conclude that $\langle L_{\rm rad}\rangle \approx \frac{1}{2}L_{\rm rad}$. At the same time, efficiency considerations impose an upper limit on $L_B$. Requiring an efficiency $\gtrsim 0.1$, then $L_B\lesssim 5 L_{\rm rad}$. We define a dimensionless parameter $C_L\equiv L_B/L_{\rm rad}$. The considerations above imply $0.5\lesssim C_L \lesssim 5$.

As shown above, equation (\ref{eq:rspher}) puts a lower limit on the emitting radius for the case of a spherical geometry. At the same time, the radius cannot be much larger, as the prompt emission radius must be smaller than the deceleration radius. Furthermore, the same expression equals (rather than just providing a limit on) the emitting radius for the case of planar geometry (see equation (\ref{eq:rplanar})). We can thus once more define a dimensionless parameter $C_R\geq 1$, which represents the emission radius in units of $2\Gamma_{\rm em}\Gamma_b c t_{\rm v}$.

Putting all of this together and making use of equations (\ref{eq:Gammaspher}), (\ref{eq:Gammapdef}), (\ref{eq:LB}) we have
\beq
\Gamma'=4.6\frac{C_R^{1/2}}{C_L^{1/4}} \bigg(\frac{\epsilon_e}{\epsilon_B}\bigg)^{-1/2}  \bigg(\frac{B_{\rm b}}{B_{\rm em}}\bigg)^{1/2} v_{g,.3c}^{-3/4}.
\label{eq:Gammaprime}
\eeq 
Magnetic reconnection models suggest that plasmoids are accelerated in parallel to the original orientation of the reconnecting field lines. This implies that the magnetic field of the emitters as seen from the bulk is not relativistically boosted as compared with the field in the emitters' frame.
Pressure balance between the plasmoids and their surroundings then implies that $B_{\rm b}\approx  B_{\rm em}$. Using this value, as well as $\epsilon_B=1, \epsilon_e=0.2$ we find $\Gamma'\approx 10 \frac{C_R^{1/2}}{C_L^{1/4}}$.

An additional constraint on $\Gamma'$ and $E$ can be obtained by considering the magnetization in the upstream of the emitting region. Since $E$ is the energy per electron in the emitters' frame, $\Gamma' E$ is the energy in the bulk frame and thus equation (\ref{Esigma}) is re-written as
\bea
E \Gamma'=0.17 \bigg(\frac{\epsilon_e}{0.2}\bigg) \sigma_{\rm up} \rm{ GeV}
\label{Esigmaplas}
\eea
Motivated by results of analytic models and PIC simulations of magnetic reconnection, we relate $\Gamma'$ to the magnetization as $\Gamma'=\sigma_{\rm up}^n$, with $0\leq n\leq 1$. For $n=0$ this parametrization reduces back to the case of no relativistic motion in the bulk frame, while reconnection models suggest that $n$ may be as large as 0.5 \citep{Lyubarsky2005}. We consider some representative values for $n$ below. We now use equation (\ref{Esigmaplas}) to relate $\Gamma'$ to $E_{\rm GeV}$ and plug the results into equations (\ref{eq:Gammaspher}), (\ref{eq:Gammapdef}). Solving for $E_{\rm GeV},\Gamma'$ as functions of $\Gamma_{\rm b}$, this leads to 
\beq
E_{\rm GeV}\!=\!(63^{6-6n}0.17^{6n} \epsilon_{e,0.2}^{7n-1}\epsilon_B^{1-n}L_{52}^{1-n}\nu_{p,5.5}^{2-2n}\Gamma_{\rm b,2.5}^{6n-6}v_{g,.3c}^{3n-3})^{1\over 4n+2},
\label{eq:EGeVsigma05}
\eeq
which reduces to
\bea
\label{eq:EGeVfinal}
& E_{\rm GeV}\!=\!6 \epsilon_{e,0.2}^{5/8}\epsilon_B^{1/8}L_{52}^{1/8}\nu_{p,5.5}^{1/4}\Gamma_{\rm b,2.5}^{-3/4}v_{g,.3c}^{-3/8} \mbox{  for }n\!=\!1/2\\ \nonumber
&E_{\rm GeV}\!=\!200 \epsilon_{e,0.2}^{1/4}\epsilon_B^{1/4}L_{52}^{1/4}\nu_{p,5.5}^{1/2}\Gamma_{\rm b,2.5}^{-3/2}v_{g,.3c}^{-3/4} \mbox{  for }n\!=\!1/4
\eea
and therefore:
\bea
\label{eq:Gammapfinal}
& \Gamma'\!=\!35 \epsilon_{e,0.2}^{-3/8}\epsilon_B^{1/8}L_{52}^{1/8}\nu_{p,5.5}^{1/4}\Gamma_{\rm b,2.5}^{-3/4}v_{g,.3c}^{-3/8} \mbox{  for }n\!=\!1/2\\ \nonumber
&\Gamma'\!=\! 10 \epsilon_{e,0.2}^{-1/4}\epsilon_B^{1/12}L_{52}^{1/12}\nu_{p,5.5}^{1/6}\Gamma_{\rm b,2.5}^{-1/2}v_{g,.3c}^{-1/4} \mbox{  for }n\!=\!1/4.
\eea
These results are consistent with the lower limits on $E_{\rm GeV}$ implied by the requirement on marginally fast solutions (\S \ref{sec:revisedlimits}) and with the upper limits implied by requirements on the IC cooling (\S \ref{sec:IC}). Interestingly, for $n=1/4$, we find $\Gamma'\approx 10$. These limits on $\Gamma'$ are consistent with the independent constraints imposed by equation (\ref{eq:Gammaprime}), as well as with expectations from relativistic turbulence or `mini-jets' models (e.g., \citealp{LB2003,Lyutikov2006,KN2009,Lazar2009,Giannios2009, zhangandzhang2014, BarniolDuran2016}) and with constraints on the variability time-scale \citep{BG2016}. These values lead to very large values of the upstream magnetization,
\bea
\label{eq:sigmafinal}
& \sigma_{\rm up}\!=\!1100 \epsilon_{e,0.2}^{-3/4}\epsilon_B^{1/4}L_{52}^{1/4}\nu_{p,5.5}^{1/2}\Gamma_{\rm b,2.5}^{-3/2}v_{g,.3c}^{-3/4} \mbox{  for }n\!=\!1/2\\ \nonumber
&\sigma_{\rm up}\!=\! 1.1\times 10^4 \epsilon_{e,0.2}^{-1}\epsilon_B^{3/4}L_{52}^{3/4}\nu_{p,5.5}^{2/3}\Gamma_{\rm b,2.5}^{-2}v_{g,.3c}^{-1} \mbox{  for }n\!=\!1/4.
\eea
We explore the implication of these large values below.

\subsection{Particle energy distribution}
\label{sec:particledist}
So far our analysis assumes that the power-law distribution of the radiating particles is such that a characteristic Lorentz factor $\gamma_{\rm e}$ dominates both in terms of the total particles' energy {\it and} number. This is a common expectation in shocks where $p>2$. However, for the extreme magnetization $\sigma_{\rm up}$ inferred for the jet (see \S \ref{sec:MatterorMag}), this assumption is likely to break down. In this Section, we explore the implications of this decoupling.

PIC simulations of reconnection find that the slope of the electrons' LF distribution, $p$, depends sensitively on  $\sigma_{\rm up}$ \citep{SironiSpitkovsky2014,Kagan2015,Guo2015,Werner2016}. These simulations find that for $\sigma_{\rm up}\gtrsim 10$, the spectra become hard, with $p<2$. For $1<p<2$, the number of electrons is dominated by the lowest LF electrons, $\gamma_{\rm min}$, while the total energy instead is dominated by the highest energy LF electrons, $\gamma_{\rm max}$. This means that the LF $\gamma_e$ associated with the energy per particle (equation (\ref{gamma_e})) is smaller than the LF of particles contributing to the peak of the emission, which for instantaneous acceleration is at $\gamma_i=\gamma_{\rm max}$. Assuming $\gamma_{\rm max}\gg \gamma_{\rm min}$, we can rewrite equation (\ref{gamma_i}) as 
\bea
&\gamma_{\rm e} m_e c^2\! =\!\frac{\int_{\gamma_{\rm min}}^{\gamma_{\rm max}}\frac{dN}{d\gamma}\gamma m_e c^2 d\gamma}{\int_{\gamma_{\rm min}}^{\gamma_{\rm max}}\frac{dN}{d\gamma}d\gamma} \! \rightarrow \! \gamma_e\!=\!\frac{p-1}{2-p}\bigg(\frac{\gamma_i}{\gamma_{\rm min}}\bigg)^{1-p}\! \gamma_i.
\eea
Assuming $\gamma_{\rm min}\approx 1$ as the most extreme case (motivated also by PIC simulations), we obtain
\beq
\gamma_i=\bigg( \frac{2-p}{p-1}\gamma_e\bigg)^{1\over 2-p}.
\label{eq:gammaip}
\eeq
As an example, for $p\approx 1.5$ (as found in simulations for $\sigma_{\rm up}\geq 50$), equation (\ref{eq:gammaip}) reduces to $\gamma_i=\gamma_e^2$. Since the energy is dominated by particles with $\gamma_i$ that are a small fraction of the total number of particles, they must be accelerated to larger energies as compared with the $p>2$ case (where $\gamma_i=\gamma_e$, see equation (\ref{gamma_i})) in order to achieve the same energy per particle. This also implies that in order to achieve a balance between heating and cooling rates, the heating rate of these particles is increased by $\gamma_i/\gamma_e$ as compared with equation (\ref{eq:heatrate}).
The result is:
\beq
\gamma_s=6.6\times 10^7 \bigg(\frac{2000(2-p)E_{\rm GeV}}{p-1}\bigg)^{1\over 2(p-2)}\frac{\nu_{\rm p,5.5}t_{\rm v,0.5}^{1/2}}{\Gamma_{\rm em,2}^{1/2}},
\eeq
which for $p=1.5$ simplifies to:
\beq
\gamma_s=3\times 10^4\frac{\nu_{\rm p,5.5}t_{\rm v,0.5}^{1/2}}{E_{\rm GeV}\Gamma_{\rm em,2}^{1/2}}.
\eeq
This leads to a decrease of the minimum allowed energy per particle ($E_{\rm tr}$) and the corresponding LF ($\gamma_{\rm tr}$) for balanced heating solutions. Their new values are (for $p=1.5$):
\bea
& E_{\rm tr}=0.2 t_{\rm v,0.5}^{1/6}\nu_{\rm p,5.5}^{1/3} \Gamma_{\rm em,2}^{-1/6}\mbox{GeV} \nonumber \\
& \gamma_{\rm tr}=1.3\times 10^5 t_{\rm v,0.5}^{1/3}\nu_{\rm p,5.5}^{2/3} \Gamma_{\rm em,2}^{-1/3}.
\label{eq:gammatrp15}
\eea
The reduced values of $\gamma_{\rm s}$ (as compared with $p>2$) lead to stronger values of the magnetic field, and therefore reduce the estimate of $\Gamma_{\rm em}$ given by equation \ref{eq:Gammaspher} (for $p=1.5$):
\beq
\Gamma_{\rm em}=1.6\times 10^3 \frac{\nu_{\rm p,5.5}^{1/3} L_{52}^{1/6}\epsilon_B^{1/6}}{\epsilon_{e,0.2}^{1/6}E_{\rm GeV}^{2/3}v_{g,.3c}^{1/2}}.
\label{eq:Gammap15}
\eeq
Combining this with $E \Gamma'\propto \sigma_{\rm up}$ (equation \ref{Esigmaplas}), $\Gamma'=\sigma_{\rm up}^n$  and equation (\ref{eq:Gammapdef}) we obtain 
\beq
E_{\rm GeV}=5^{3(1-n)\over n+2}0.17^{3n\over n+2}\epsilon_{e,0.2}^{7n-1 \over 2(n+2)}\epsilon_B^{1-n\over 2n+4}\Gamma_{\rm b,2.5}^{-3(1-n)\over n+2}\nu_{\rm p,5.5}^{1-n\over n+2}L_{52}^{1-n \over 2(n+2)}v_{g,.3c}^{3n-3 \over 2n+4},
\eeq
which reduces to
\bea
\label{eq:EGeVp15}
& E_{\rm GeV}\!=\!0.9 \epsilon_{e,0.2}^{1/2}\epsilon_B^{1/10} L_{52}^{1/10}\nu_{p,5.5}^{1/5}\Gamma_{\rm b,2.5}^{-3/5}v_{g,.3c}^{-3/10} \mbox{  for }n\!=\!1/2 \nonumber \\
&E_{\rm GeV}\!=\!2.7 \epsilon_{e,0.2}^{1/6}\epsilon_B^{1/6}L_{52}^{1/6}\nu_{p,5.5}^{1/3}\Gamma_{\rm b,2.5}^{-1}v_{g,.3c}^{-1/2} \mbox{  for }n\!=\!1/4
\eea
and:
\bea
\label{eq:Gammapp15}
& \Gamma'\!=\!5 \epsilon_{e,0.2}^{-0.5}\epsilon_B^{1/10}L_{52}^{1/10}\nu_{p,5.5}^{1/5}\Gamma_{\rm b,2.5}^{-3/5}v_{g,.3c}^{-3/10} \mbox{  for }n\!=\!1/2\\ \nonumber
&\Gamma'\!=\! 2.5 \epsilon_{e,0.2}^{-0.28}\epsilon_B^{1/18}L_{52}^{1/18}\nu_{p,5.5}^{1/9}\Gamma_{\rm b,2.5}^{-1/3}v_{g,.3c}^{-1/6} \mbox{  for }n\!=\!1/4.
\eea
These values are consistent with the condition $E_{\rm GeV}\geq E_{\rm tr}$. In addition, they demonstrate that even for $p\approx 1.5$, the plasmoids are expected to be moving at least mildly relativistically in the bulk frame. It is interesting to note that these values of $E,\Gamma',n$ correspond to $31 \lesssim \sigma_{\rm up}\lesssim 45$ which is indeed consistent with the value of $p\approx 1.5$ assumed here. One concern with this scenario however, is that due to the increased energy density implied for a given $E$ as compared with the case of $p>2$, the required $E$ in order for IC cooling to be sub-dominant to synchrotron (equation (\ref{eq:upperE})) is reduced.
Following the same procedure described in \S \ref{sec:IC} for $p>2$ (but using the revised values for $\gamma_s$, $\Gamma_{\rm em}$ appropriate for $p=1.5$) we find
\beq
E<2 \frac{\epsilon_B^{21/96}\nu_{\rm p,5.5}^{51/96}t_{\rm v,0.5}^{7/16}v_{g,.3c}^{3/32}}{\epsilon_{e,0.2}^{21/96}L_{52}^{5/32}}\mbox{GeV}.
\eeq
These values are considerably smaller than the equivalent limits for $p>2$ although still consistent with the energies implied by equation (\ref{eq:EGeVp15}), assuming $n\approx 1/2$.

An additional limit on the electrons' LFs, $\gamma_s$, is obtained by equating the Larmour acceleration time with the energy loss time due to synchrotron \citep{deJager1996}\footnote{A related limit arises from requiring that the size of the emitting region must be smaller than the Larmour radius of the highest energy particles. This consideration results in the same scaling for $\gamma_{\rm L}$, but reduced by a factor $(v_g/c)^{1/2}\approx 0.5$, and does not change the qualitative conclusion below.}:
\beq
\gamma_{\rm L}\approx 4\times 10^7B_{\rm em}^{-1/2}\approx 2.6\times 10^6 t_{\rm v,0.5}^{1/2}\nu_{\rm p,5.5}^{1/2} E_{\rm GeV}^{-1}.
\eeq
Since these values are more than an order of magnitude above $\gamma_{\rm tr}$ (given by equation (\ref{eq:gammatrp15})) and since balanced heating solutions have $\gamma_s<\gamma_{\rm tr}$, the Larmour limits are consistent with the picture presented here.

\section{Discussion}
\label{sec:Dis}
\subsection{Shorter heating times}
\label{sec:shortheat}
Consider a variant on the slow heating model presented in this paper where each particle undergoes heating for only a fraction $\alpha\leq 1$ of $t_{\rm v}$. This scenario implies that particles experience a balance between heating and cooling for a time $\alpha t_{\rm v}$, after which acceleration ceases and they spend the rest of the dynamical time in fast cooling conditions. Even though the particles maintain a balance between heating and cooling for only a small part of the dynamical time, the overall spectrum emitted by those particles will resemble a slow cooling slope rather than a fast cooling one. The reason is that the overall energy emitted by the particles is $E$, while at the end of their heating they have a LF $\gamma_{\rm s}$ and, as can be seen by figure \ref{fig1}, for slow heating solutions $\gamma_{\rm s}m_ec^2\ll E$. Thus, although the particles may spend a large amount of time in fast cooling conditions, only a small portion of their total emitted energy is released at this stage. These shorter lived `balanced heating' conditions can therefore satisfy the requirement on the low-energy spectral slope while maintaining a large efficiency. It is thus interesting to consider how it can affect the results presented in this paper. 

In order to obtain the same energy per particle, $E$, the heating rate, $\dot{\epsilon_{\rm h}}$, is increased by $\alpha^{-1}$ as compared with equation (\ref{eq:heatrate}), while the cooling rate given by equation (\ref{eq:coolrate}) remains the same. The result is that $\gamma_{\rm s}$ is reduced by $\alpha^{1/2}$ and $E_{\rm tr},\gamma_{\rm tr}$ are both reduced by $\alpha^{1/3}$ as compared with their values for the slow heating case given by equations (\ref{eq:Etr}) and (\ref{eq:gtr}). Faster heating also reduces somewhat the constraint on $\Gamma_{\rm em}$ obtained by equation (\ref{eq:Gammaspher}): $\Gamma_{\rm em}\propto B_{\rm em}(\gamma_{\rm s})^{-1/3}\propto \gamma_{\rm s}^{-2/3}\propto \alpha^{1/3}$. At face value, it would seem that $\alpha \ll 1$ could thus mitigate the requirement on relativistic motion in the bulk frame. However, solutions with $\alpha \ll 1$ require very strong magnetic fields $B(\gamma_s)\propto \alpha^{-1}$, and as a result correspond to a huge Poynting luminosity. Assuming $B_{\rm b}=B_{\rm em}$ (see \ref{sec:relmotion}) and taking the minimum allowed radius as implied by variability (equation (\ref{eq:rspher}); results become more constraining for larger radii) we obtain that $L_B\propto R^2 B^2 \propto \alpha^{-4/3}$ while $L_{\rm rad}$ is unchanged. Therefore, balanced heating with small values of $\alpha$ will lead to extremely inefficient bursts with $L_B \gg L_{\rm rad}$. Assuming $L_B<5 L_{\rm rad}$, we find that $\alpha \gtrsim 0.3$. We conclude that balanced heating solutions require a heating time that is not much smaller than the dynamical time. 

\subsection{Spectral shape}
\label{sec:specshape}
We have discussed in this work the characteristic energies of the particles that contribute to the peak of the $\gamma$-ray emission regardless of the specific particle acceleration mechanism. The considerations that we made here are generic requirements on a synchrotron signal such that it would not result in a strongly fast cooling spectrum (which will be in strong contradiction with observations). The exact shape of the spectrum however, could still be affected by the details of the particle acceleration mechanism. The effects of slow heating acceleration mechanisms on the particle spectrum have been discussed in the literature by various authors (e.g., \citealt{asano2009,Brunetti2016,Xu2017}). We refer the reader to those papers for a more in depth discussion of how acceleration can modify the particle spectrum.

A major consideration regarding the acceleration mechanism, has to do with the rate at which particles of different energies are energized. Since the energy loss rate via synchrotron (as well as in IC if the Klein-Nishina effect can be neglected) scales as $P_{\rm cool}\propto \gamma^2$, this implies that in order for energy balance to hold for particles of all energies, one should also have $P_{\rm heat}\propto \gamma^2$. This condition may be difficult to obtain in practice. This is because in Fermi type II acceleration, the energy gain rate scales as $P_{\rm heat}\propto D/\gamma$, where $D$ is the diffusion coefficient. $D$ is then expected to scale as $\gamma^n$ with $n=2$ for small scale MHD turbulence, $n=5/3$ for Kolmogorov turbulence, or at the limit of fastest acceleration, or $n=1$ for strong turbulence (also known as the Bohm limit). The energy gain rate is therefore expected to be constant or decreasing with $\gamma$ and either way is softer (as a function of $\gamma$) than $P_{\rm cool}$. The implication is that if energy balance is maintained for particles with $\gamma\approx \gamma_s$, then particles with an initial LF $\gamma<\gamma_{\rm s}$ will heat faster than they cool, while particles with $\gamma>\gamma_{\rm s}$ cool faster than they heat. This would lead to an eventual bunching up of particles around $\gamma_s$.
Furthermore, this demonstrates that in order to initially accelerate particles to $\gamma>\gamma_{\rm s}$, there must, in fact be two distinct acceleration processes taking place. The first, creating the $dN/d\gamma\propto \gamma^{-p}$ distribution on a short time-scale, and the second, slowly heating the electrons such that cooling balances acceleration at $\gamma=\gamma_{\rm s}.$ If the initial acceleration process is not present, no particles will reach $\gamma>\gamma_{\rm s}$, and the spectrum would cut-off sharply beyond the sub-MeV peak, contrary to observations.

As particles above $\gamma_{\rm s}$ (assuming that such particles exist, i.e., that $p>2$) are essentially in fast cooling conditions, their cooling would result in a spectrum $F_{\nu}\propto \nu^{-p/2}$ for $\nu>\nu_{\rm syn}(\gamma_s)$ (where $p$ is the slope of the initial particle spectra). This is similar to the spectrum from instantaneously accelerated electrons radiating at $\nu>\max(\nu_m,\nu_c)$. Particles with $\gamma\ll \gamma_{\rm s}$ will heat up to $\gamma_{\rm s}$ over a dynamical time-scale. Since the heating rate is a decreasing (or at most flat) function of the electrons LF, electrons starting at $\gamma\ll \gamma_{\rm s}$ will spend a short time (compared to $t_{\rm v}$) at their initial LFs. As a result most of the emission from these electrons will take place after they reach $\gamma\approx \gamma_{\rm s}$. Therefore, at the synchrotron frequencies corresponding to $\gamma\approx \gamma_{\rm s}$, the spectrum will be dominated by the classical $F_{\nu}\propto \nu^{1/3}$ contribution of electrons at $\gamma\approx \gamma_{\rm s}$. As $\gamma$ approaches $\gamma_{\rm s}$ the ratio of the heating to the dynamical time becomes closer to the unity, and the emitted spectrum becomes slightly softer than $\nu^{1/3}$. The softness of the spectrum in this range is always limited however by $F_{\nu}\propto \nu^{1-p\over 2}$ which is the slow cooling spectrum emitted by a stationary (unheated) distribution of the type $dN/d\gamma \propto \gamma^{-p}$. 

\subsection{Comparison to other studies}
\label{sec:compare}
We have focused in this work on the required conditions at the emitting region needed to account for marginally fast cooling of the $\gamma$-ray emitting electrons via synchrotron. This topic has been studied in the literature by different authors (e.g., \citealt{Ghisellini1999,pawananderin,Daigne2011,pazandtsvi,BP2014}).
These studies can be divided into three groups. First, e.g., \cite{Ghisellini1999,Peer2006,Giannios2008} discussed continuous heating in the context of photospheric models, where the electrons are sub or at most mildly relativistic. At these conditions, synchrotron photons are self absorbed and the emission is dominated by multiple IC scatterings of the synchrotron seed. The second group of studies consider general synchrotron models \citep{pawananderin,pazandtsvi,BP2014}. In these, if the acceleration is instantaneous, the resulting parameter space is characterized by a large radius ($R\approx 6\times 10^{16}$cm), a large Lorentz factor of the emitting material ($\Gamma_{\rm em}\approx2000$), a large electrons' LF ($\gamma_e\approx 10^5$), and weak magnetic fields ($B_{\rm em}\approx$ few $G$). Introducing re-acceleration, increases the required magnetic field and therefore reduces somewhat $\gamma_e$. These results are consistent with the parameter ranges found in the current study. Furthermore, since the required number of electrons is significantly decreased in this case (since it scales as $f^{-1}$ and $f\gg1$), the implied energy per electron associated with these solutions is larger than for the instantaneous case. This, as well, is consistent with the findings reported here (although the formulation of the problem was quite different in the earlier studies). %A major difference in those previous works and the current paper, is that since they considered multiple, distinct, re-acceleration processes, it was not clear what sets the number of acceleration episodes that would allow for marginally fast cooling. Alternatively, in the current work imposing continuous heating, naturally leads 
The third group of studies \citep{Daigne2011} and section 4 of \cite{pazandtsvi}, also considered synchrotron solutions, but specifically in the context of internal shocks. In this case, since the jet is baryonic and the energy per electrons is $\lesssim 0.2$GeV (same as equation (\ref{Esigma}) but for $\sigma_{\rm up}\rightarrow1$), slow heating conditions are not possible unless only a small fraction of the electrons $\xi \lesssim 10^{-2}$ are accelerated to relativistic energies by the shock. Since these conditions are not supported by PIC simulations of acceleration in shocks which demonstrate that practically all electrons undergo heating behind the shock front \citep{SKL2015} (as well as in reconnection, see \citealt{Sironi2015}), we have assumed that $\xi=1$ in the current work.

\section{Conclusions}
\label{sec:conclusion}
The synchrotron mechanism has been widely discussed for the prompt phase of GRBs. Previous studies have shown that the physical conditions at the emitting region typically lead to the electrons cooling via synchrotron on a very short time-scale (as compared with the dynamical one). This results in a low-energy spectral slope that is in strong contention with observations. This problem may be overcome if the electrons' energy losses due to synchrotron are balanced by a continuous source of heating, leading to `marginally fast cooling' electrons ($\nu_c \approx \nu_m$). Here we revisit the model and derive some general constrains on any synchrotron model based on basic observed properties of GRBs: the characteristic sub-MeV energy where the emission peaks, the hardness of the slope below the peak, the characteristic luminosity and the variability time-scale of GRBs.

If the peak emission is dominated by the majority of the particles (as expected in shock models or low $\sigma$ reconnection), the emitting region has to be characterized by $\Gamma_{\rm em} \gtrsim 3000$, well in excess of what is inferred for the bulk jet motion from afterglow modeling. Several independent constraints indicate that emitters have to be characterized by fast motion ($\Gamma'\sim10$) in the rest frame of the jet and that, at the emitting region, the jet must be in the high-$\sigma$ regime (where the energy per electron can reach $E\gtrsim20$GeV). In such a regime, the particle distribution is hard so that most of the energy is injected in a minority of the particles and the constraints on the bulk motions are somewhat relaxed. Synchrotron-only models work for $R>10^{16}$cm, $\sigma\sim 30-50$, $\Gamma'\sim$ several.

These results can be used as a basis for future PIC simulations of magnetic reconnection in GRBs. Since simulations are extremely computationally demanding, this work may prove to be critical to narrow down the possible parameter space that could lead to the observed properties of GRBs.

\section*{Acknowledgements}

We thank Jonathan Granot, Pawan Kumar, Maxim Lyutikov, Lara Nava and Tsvi Piran for useful discussions and comments.
DG acknowledges support from NASA through the grants NNX16AB32G and NNX17AG21G issued through the Astrophysics Theory Program. 

%%%%%%%%%%%%%%%%%%%%%%%%%%%%%%%%%%%%%%%%%%%%%%%%%%



%\bibliographystyle{mnras}

\begin{thebibliography}{85}
	\expandafter\ifx\csname natexlab\endcsname\relax\def\natexlab#1{#1}\fi
	
	\bibitem[{{Abbott} {et~al}\mbox{.}(2017){Abbott}, {Abbott}, {Abbott},
		{Acernese}, {Ackley}, {Adams}, {Adams}, {Addesso}, {Adhikari}, {Adya}, \&
		et~al.}]{abbott2017}
	{Abbott} B.~P. {et~al.}, 2017, \apjl, 848, L13
	
	\bibitem[{{Abramowicz}, {Novikov} \& {Paczynski}(1991){Abramowicz}, {Novikov},
		\& {Paczynski}}]{Abramowicz1991}
	{Abramowicz} M.~A., {Novikov} I.~D., {Paczynski} B., 1991, \apj, 369, 175
	
	\bibitem[{{Achterberg} {et~al}\mbox{.}(2001){Achterberg}, {Gallant}, {Kirk}, \&
		{Guthmann}}]{Achterberg2001}
	{Achterberg} A., {Gallant} Y.~A., {Kirk} J.~G., {Guthmann} A.~W., 2001, \mnras,
	328, 393
	
	\bibitem[{{Ando}, {Nakar} \& {Sari}(2008){Ando}, {Nakar}, \& {Sari}}]{Ando2008}
	{Ando} S., {Nakar} E., {Sari} R., 2008, \apj, 689, 1150
	
	\bibitem[{{Asano} \& {Terasawa}(2009)}]{asano2009}
	{Asano} K., {Terasawa} T., 2009, \apj, 705, 1714
	
	\bibitem[{{Barniol Duran} \& {Kumar}(2009)}]{BarniolDuran2009}
	{Barniol Duran} R., {Kumar} P., 2009, \mnras, 395, 955
	
	\bibitem[{{Barniol Duran}, {Leng} \& {Giannios}(2016){Barniol Duran}, {Leng},
		\& {Giannios}}]{BarniolDuran2016}
	{Barniol Duran} R., {Leng} M., {Giannios} D., 2016, \mnras, 455, L6
	
	\bibitem[{{Bednarz} \& {Ostrowski}(1998)}]{Bednarz1998}
	{Bednarz} J., {Ostrowski} M., 1998, Physical Review Letters, 80, 3911
	
	\bibitem[{{Beloborodov}(2010)}]{Beloborodov2010}
	{Beloborodov} A.~M., 2010, \mnras, 407, 1033
	
	\bibitem[{{Beniamini} \& {Giannios}(2017)}]{BG2017}
	{Beniamini} P., {Giannios} D., 2017, \mnras, 468, 3202
	
	\bibitem[{{Beniamini}, {Giannios} \& {Metzger}(2017){Beniamini}, {Giannios}, \&
		{Metzger}}]{BGM2017}
	{Beniamini} P., {Giannios} D., {Metzger} B.~D., 2017, \mnras, 472, 3058
	
	\bibitem[{{Beniamini} \& {Granot}(2016)}]{BG2016}
	{Beniamini} P., {Granot} J., 2016, \mnras, 459, 3635
	
	\bibitem[{{Beniamini} {et~al}\mbox{.}(2015){Beniamini}, {Nava}, {Duran}, \&
		{Piran}}]{Beniamini2015}
	{Beniamini} P., {Nava} L., {Duran} R.~B., {Piran} T., 2015, \mnras, 454, 1073
	
	\bibitem[{{Beniamini}, {Nava} \& {Piran}(2016){Beniamini}, {Nava}, \&
		{Piran}}]{Beniamini2016}
	{Beniamini} P., {Nava} L., {Piran} T., 2016, \mnras, 461, 51
	
	\bibitem[{{Beniamini} \& {Piran}(2013)}]{pazandtsvi}
	{Beniamini} P., {Piran} T., 2013, \apj, 769, 69
	
	\bibitem[{{Beniamini} \& {Piran}(2014)}]{BP2014}
	{Beniamini} P., {Piran} T., 2014, \mnras, 445, 3892
	
	\bibitem[{{Brunetti} \& {Lazarian}(2016)}]{Brunetti2016}
	{Brunetti} G., {Lazarian} A., 2016, \mnras, 458, 2584
	
	\bibitem[{{Burgess}(2017)}]{Burgess2017}
	{Burgess} J.~M., 2017, ArXiv e-prints
	
	\bibitem[{{Cerutti} {et~al}\mbox{.}(2012){Cerutti}, {Werner}, {Uzdensky}, \&
		{Begelman}}]{Cerutti2012}
	{Cerutti} B., {Werner} G.~R., {Uzdensky} D.~A., {Begelman} M.~C., 2012, \apjl,
	754, L33
	
	\bibitem[{{Daigne}, {Bo{\v s}njak} \& {Dubus}(2011){Daigne}, {Bo{\v s}njak}, \&
		{Dubus}}]{Daigne2011}
	{Daigne} F., {Bo{\v s}njak} {\v Z}., {Dubus} G., 2011, \aap, 526, A110
	
	\bibitem[{{de Jager} {et~al}\mbox{.}(1996){de Jager}, {Harding}, {Michelson},
		{Nel}, {Nolan}, {Sreekumar}, \& {Thompson}}]{deJager1996}
	{de Jager} O.~C., {Harding} A.~K., {Michelson} P.~F., {Nel} H.~I., {Nolan}
	P.~L., {Sreekumar} P., {Thompson} D.~J., 1996, \apj, 457, 253
	
	\bibitem[{{Fan} \& {Piran}(2006)}]{Fan2006}
	{Fan} Y., {Piran} T., 2006, \mnras, 369, 197
	
	\bibitem[{{Fan}(2010)}]{Fan2010}
	{Fan} Y.-Z., 2010, \mnras, 403, 483
	
	\bibitem[{{Fan} \& {Wei}(2005)}]{Fan2005}
	{Fan} Y.~Z., {Wei} D.~M., 2005, \mnras, 364, L42
	
	\bibitem[{{Fenimore}, {Epstein} \& {Ho}(1993){Fenimore}, {Epstein}, \&
		{Ho}}]{Fenimore1993}
	{Fenimore} E.~E., {Epstein} R.~I., {Ho} C., 1993, \aaps, 97, 59
	
	\bibitem[{{Fishman} \& {Meegan}(1995)}]{Fishman1995}
	{Fishman} G.~J., {Meegan} C.~A., 1995, \araa, 33, 415
	
	\bibitem[{{Ghirlanda}, {Celotti} \& {Ghisellini}(2002){Ghirlanda}, {Celotti},
		\& {Ghisellini}}]{Ghirlanda2002}
	{Ghirlanda} G., {Celotti} A., {Ghisellini} G., 2002, \aap, 393, 409
	
	\bibitem[{{Ghirlanda} {et~al}\mbox{.}(2012){Ghirlanda}, {Nava}, {Ghisellini},
		{Celotti}, {Burlon}, {Covino}, \& {Melandri}}]{Ghirlanda2012}
	{Ghirlanda} G., {Nava} L., {Ghisellini} G., {Celotti} A., {Burlon} D., {Covino}
	S., {Melandri} A., 2012, \mnras, 420, 483
	
	\bibitem[{{Ghisellini} \& {Celotti}(1999)}]{Ghisellini1999}
	{Ghisellini} G., {Celotti} A., 1999, \apjl, 511, L93
	
	\bibitem[{{Giannios}(2006)}]{Giannios2006}
	{Giannios} D., 2006, \aap, 457, 763
	
	\bibitem[{{Giannios}(2008)}]{Giannios2008}
	{Giannios} D., 2008, \aap, 480, 305
	
	\bibitem[{{Giannios}(2012)}]{Giannios2012}
	{Giannios} D., 2012, \mnras, 422, 3092
	
	\bibitem[{{Giannios}(2013)}]{Giannios2013}
	{Giannios} D., 2013, \mnras, 431, 355
	
	\bibitem[{{Giannios} \& {Spruit}(2005)}]{gianniosandspruit2005}
	{Giannios} D., {Spruit} H.~C., 2005, \aap, 430, 1
	
	\bibitem[{{Giannios}, {Uzdensky} \& {Begelman}(2009){Giannios}, {Uzdensky}, \&
		{Begelman}}]{Giannios2009}
	{Giannios} D., {Uzdensky} D.~A., {Begelman} M.~C., 2009, \mnras, 395, L29
	
	\bibitem[{{Goodman}(1986)}]{Goodman1986}
	{Goodman} J., 1986, \apjl, 308, L47
	
	\bibitem[{{Guiriec} {et~al}\mbox{.}(2015){Guiriec}, {Kouveliotou}, {Daigne},
		{Zhang}, {Hasco{\"e}t}, {Nemmen}, {Thompson}, {Bhat}, {Gehrels}, {Gonzalez},
		{Kaneko}, {McEnery}, {Mochkovitch}, {Racusin}, {Ryde}, {Sacahui}, \&
		{{\"U}nsal}}]{Guiriec2015}
	{Guiriec} S. {et~al.}, 2015, \apj, 807, 148
	
	\bibitem[{{Guo} {et~al}\mbox{.}(2014){Guo}, {Li}, {Daughton}, \&
		{Liu}}]{Guo2014}
	{Guo} F., {Li} H., {Daughton} W., {Liu} Y.-H., 2014, Physical Review Letters,
	113, 155005
	
	\bibitem[{{Guo} {et~al}\mbox{.}(2015){Guo}, {Liu}, {Daughton}, \&
		{Li}}]{Guo2015}
	{Guo} F., {Liu} Y.-H., {Daughton} W., {Li} H., 2015, \apj, 806, 167
	
	\bibitem[{{Hakkila} \& {Preece}(2011)}]{Hakkila2011}
	{Hakkila} J., {Preece} R.~D., 2011, \apj, 740, 104
	
	\bibitem[{{Heavens} \& {Drury}(1988)}]{Heavens1988}
	{Heavens} A.~F., {Drury} L.~O., 1988, \mnras, 235, 997
	
	\bibitem[{{Kagan} {et~al}\mbox{.}(2015){Kagan}, {Sironi}, {Cerutti}, \&
		{Giannios}}]{Kagan2015}
	{Kagan} D., {Sironi} L., {Cerutti} B., {Giannios} D., 2015, \ssr, 191, 545
	
	\bibitem[{{Kaneko} {et~al}\mbox{.}(2006){Kaneko}, {Preece}, {Briggs},
		{Paciesas}, {Meegan}, \& {Band}}]{Kaneko2006}
	{Kaneko} Y., {Preece} R.~D., {Briggs} M.~S., {Paciesas} W.~S., {Meegan} C.~A.,
	{Band} D.~L., 2006, \apjs, 166, 298
	
	\bibitem[{{Katz}(1994)}]{katz1994}
	{Katz} J.~I., 1994, \apjl, 432, L107
	
	\bibitem[{{Kumar} \& {McMahon}(2008)}]{pawananderin}
	{Kumar} P., {McMahon} E., 2008, \mnras, 384, 33
	
	\bibitem[{{Kumar} \& {Narayan}(2009)}]{KN2009}
	{Kumar} P., {Narayan} R., 2009, \mnras, 395, 472
	
	\bibitem[{{Kumar} \& {Panaitescu}(2000)}]{KP2000}
	{Kumar} P., {Panaitescu} A., 2000, \apjl, 541, L51
	
	\bibitem[{{Kumar} \& {Zhang}(2015)}]{kumarandzhang}
	{Kumar} P., {Zhang} B., 2015, \physrep, 561, 1
	
	\bibitem[{{Lazar}, {Nakar} \& {Piran}(2009){Lazar}, {Nakar}, \&
		{Piran}}]{Lazar2009}
	{Lazar} A., {Nakar} E., {Piran} T., 2009, \apjl, 695, L10
	
	\bibitem[{{Lazzati} \& {Begelman}(2010)}]{Lazzati2010}
	{Lazzati} D., {Begelman} M.~C., 2010, \apj, 725, 1137
	
	\bibitem[{{Liang} {et~al}\mbox{.}(2010){Liang}, {Yi}, {Zhang}, {L{\"u}},
		{Zhang}, \& {Zhang}}]{Liang2010}
	{Liang} E.-W., {Yi} S.-X., {Zhang} J., {L{\"u}} H.-J., {Zhang} B.-B., {Zhang}
	B., 2010, \apj, 725, 2209
	
	\bibitem[{{Lithwick} \& {Sari}(2001)}]{Lithwick2001}
	{Lithwick} Y., {Sari} R., 2001, \apj, 555, 540
	
	\bibitem[{{Liu} {et~al}\mbox{.}(2015){Liu}, {Guo}, {Daughton}, {Li}, \&
		{Hesse}}]{Liu2015}
	{Liu} Y.-H., {Guo} F., {Daughton} W., {Li} H., {Hesse} M., 2015, Physical
	Review Letters, 114, 095002
	
	\bibitem[{{L{\"u}} {et~al}\mbox{.}(2012){L{\"u}}, {Zou}, {Lei}, {Zhang}, {Wu},
		{Wang}, {Liang}, \& {L{\"u}}}]{Lu2012}
	{L{\"u}} J., {Zou} Y.-C., {Lei} W.-H., {Zhang} B., {Wu} Q., {Wang} D.-X.,
	{Liang} E.-W., {L{\"u}} H.-J., 2012, \apj, 751, 49
	
	\bibitem[{{Lyubarsky}(2005)}]{Lyubarsky2005}
	{Lyubarsky} Y.~E., 2005, \mnras, 358, 113
	
	\bibitem[{{Lyutikov}(2006)}]{Lyutikov2006}
	{Lyutikov} M., 2006, MNRAS, 369, L5
	
	\bibitem[{{Lyutikov} \& {Blandford}(2003)}]{LB2003}
	{Lyutikov} M., {Blandford} R., 2003, ArXiv Astrophysics e-prints
	
	\bibitem[{{Melzani} {et~al}\mbox{.}(2014){Melzani}, {Walder}, {Folini},
		{Winisdoerffer}, \& {Favre}}]{Melzani2014}
	{Melzani} M., {Walder} R., {Folini} D., {Winisdoerffer} C., {Favre} J.~M.,
	2014, \aap, 570, A112
	
	\bibitem[{{M{\'e}sz{\'a}ros} \& {Rees}(2000)}]{Meszaros2000}
	{M{\'e}sz{\'a}ros} P., {Rees} M.~J., 2000, \apj, 530, 292
	
	\bibitem[{{Nakar} \& {Piran}(2002)}]{NP2002}
	{Nakar} E., {Piran} T., 2002, \mnras, 331, 40
	
	\bibitem[{{Nava} {et~al}\mbox{.}(2011){Nava}, {Ghirlanda}, {Ghisellini}, \&
		{Celotti}}]{Nava2011}
	{Nava} L., {Ghirlanda} G., {Ghisellini} G., {Celotti} A., 2011, \aap, 530, A21
	
	\bibitem[{{Nemiroff} {et~al}\mbox{.}(1994){Nemiroff}, {Norris}, {Kouveliotou},
		{Fishman}, {Meegan}, \& {Paciesas}}]{Nemiroff1994}
	{Nemiroff} R.~J., {Norris} J.~P., {Kouveliotou} C., {Fishman} G.~J., {Meegan}
	C.~A., {Paciesas} W.~S., 1994, \apj, 423, 432
	
	\bibitem[{{Norris} {et~al}\mbox{.}(1996){Norris}, {Nemiroff}, {Bonnell},
		{Scargle}, {Kouveliotou}, {Paciesas}, {Meegan}, \& {Fishman}}]{Norris1996}
	{Norris} J.~P., {Nemiroff} R.~J., {Bonnell} J.~T., {Scargle} J.~D.,
	{Kouveliotou} C., {Paciesas} W.~S., {Meegan} C.~A., {Fishman} G.~J., 1996,
	\apj, 459, 393
	
	\bibitem[{{Oganesyan} {et~al}\mbox{.}(2017{\natexlab{a}}){Oganesyan}, {Nava},
		{Ghirlanda}, \& {Celotti}}]{Oganesyan2017b}
	{Oganesyan} G., {Nava} L., {Ghirlanda} G., {Celotti} A., 2017{\natexlab{a}},
	ArXiv e-prints
	
	\bibitem[{{Oganesyan} {et~al}\mbox{.}(2017{\natexlab{b}}){Oganesyan}, {Nava},
		{Ghirlanda}, \& {Celotti}}]{Oganesyan2017a}
	{Oganesyan} G., {Nava} L., {Ghirlanda} G., {Celotti} A., 2017{\natexlab{b}},
	\apj, 846, 137
	
	\bibitem[{{Pe'er}(2015)}]{Pe'er2015}
	{Pe'er} A., 2015, Advances in Astronomy, 2015, 907321
	
	\bibitem[{{Pe'er}, {M{\'e}sz{\'a}ros} \& {Rees}(2006){Pe'er},
		{M{\'e}sz{\'a}ros}, \& {Rees}}]{Peer2006}
	{Pe'er} A., {M{\'e}sz{\'a}ros} P., {Rees} M.~J., 2006, \apj, 642, 995
	
	\bibitem[{{Pe'er} {et~al}\mbox{.}(2012){Pe'er}, {Zhang}, {Ryde}, {McGlynn},
		{Zhang}, {Preece}, \& {Kouveliotou}}]{Pe'er2012}
	{Pe'er} A., {Zhang} B.-B., {Ryde} F., {McGlynn} S., {Zhang} B., {Preece} R.~D.,
	{Kouveliotou} C., 2012, \mnras, 420, 468
	
	\bibitem[{{Petropoulou}, {Giannios} \& {Sironi}(2016){Petropoulou}, {Giannios},
		\& {Sironi}}]{mariaetal2016}
	{Petropoulou} M., {Giannios} D., {Sironi} L., 2016, \mnras, 462, 3325
	
	\bibitem[{{Preece} {et~al}\mbox{.}(1998){Preece}, {Briggs}, {Mallozzi},
		{Pendleton}, {Paciesas}, \& {Band}}]{Preece1998}
	{Preece} R.~D., {Briggs} M.~S., {Mallozzi} R.~S., {Pendleton} G.~N., {Paciesas}
	W.~S., {Band} D.~L., 1998, \apjl, 506, L23
	
	\bibitem[{{Preece} {et~al}\mbox{.}(2000){Preece}, {Briggs}, {Mallozzi},
		{Pendleton}, {Paciesas}, \& {Band}}]{Preece2000}
	{Preece} R.~D., {Briggs} M.~S., {Mallozzi} R.~S., {Pendleton} G.~N., {Paciesas}
	W.~S., {Band} D.~L., 2000, \apjs, 126, 19
	
	\bibitem[{{Quilligan} {et~al}\mbox{.}(2002){Quilligan}, {McBreen}, {Hanlon},
		{McBreen}, {Hurley}, \& {Watson}}]{Quilligan2002}
	{Quilligan} F., {McBreen} B., {Hanlon} L., {McBreen} S., {Hurley} K.~J.,
	{Watson} D., 2002, \aap, 385, 377
	
	\bibitem[{{Ravasio} {et~al}\mbox{.}(2017){Ravasio}, {Oganesyan}, {Ghirlanda},
		{Nava}, {Ghisellini}, {Pescalli}, \& {Celotti}}]{2017Ravasio}
	{Ravasio} M.~E., {Oganesyan} G., {Ghirlanda} G., {Nava} L., {Ghisellini} G.,
	{Pescalli} A., {Celotti} A., 2017, ArXiv e-prints
	
	\bibitem[{{Rees} \& {Meszaros}(1994)}]{reesandmeszaros1994}
	{Rees} M.~J., {Meszaros} P., 1994, \apjl, 430, L93
	
	\bibitem[{{Ryde} {et~al}\mbox{.}(2010){Ryde}, {Axelsson}, {Zhang}, {McGlynn},
		{Pe'er}, {Lundman}, {Larsson}, {Battelino}, {Zhang}, {Bissaldi}, {Bregeon},
		{Briggs}, {Chiang}, {de Palma}, {Guiriec}, {Larsson}, {Longo}, {McBreen},
		{Omodei}, {Petrosian}, {Preece}, \& {van der Horst}}]{Ryde2010}
	{Ryde} F. {et~al.}, 2010, \apjl, 709, L172
	
	\bibitem[{{Sari}, {Narayan} \& {Piran}(1996){Sari}, {Narayan}, \&
		{Piran}}]{sarietal1996}
	{Sari} R., {Narayan} R., {Piran} T., 1996, \apj, 473, 204
	
	\bibitem[{{Sironi}, {Keshet} \& {Lemoine}(2015){Sironi}, {Keshet}, \&
		{Lemoine}}]{SKL2015}
	{Sironi} L., {Keshet} U., {Lemoine} M., 2015, \ssr, 191, 519
	
	\bibitem[{{Sironi}, {Petropoulou} \& {Giannios}(2015){Sironi}, {Petropoulou},
		\& {Giannios}}]{Sironi2015}
	{Sironi} L., {Petropoulou} M., {Giannios} D., 2015, \mnras, 450, 183
	
	\bibitem[{{Sironi} \& {Spitkovsky}(2014)}]{SironiSpitkovsky2014}
	{Sironi} L., {Spitkovsky} A., 2014, \apjl, 783, L21
	
	\bibitem[{{Tagliaferri} {et~al}\mbox{.}(2005){Tagliaferri}, {Goad},
		{Chincarini}, {Moretti}, {Campana}, {Burrows}, {Perri}, {Barthelmy},
		{Gehrels}, {Krimm}, {Sakamoto}, {Kumar}, {M{\'e}sz{\'a}ros}, {Kobayashi},
		{Zhang}, {Angelini}, {Banat}, {Beardmore}, {Capalbi}, {Covino}, {Cusumano},
		{Giommi}, {Godet}, {Hill}, {Kennea}, {Mangano}, {Morris}, {Nousek},
		{O'Brien}, {Osborne}, {Pagani}, {Page}, {Romano}, {Stella}, \&
		{Wells}}]{Tagliaferri2005}
	{Tagliaferri} G. {et~al.}, 2005, \nat, 436, 985
	
	\bibitem[{{Thompson}(1994)}]{Thompson1994}
	{Thompson} C., 1994, \mnras, 270, 480
	
	\bibitem[{{Werner} {et~al}\mbox{.}(2016){Werner}, {Uzdensky}, {Cerutti},
		{Nalewajko}, \& {Begelman}}]{Werner2016}
	{Werner} G.~R., {Uzdensky} D.~A., {Cerutti} B., {Nalewajko} K., {Begelman}
	M.~C., 2016, \apjl, 816, L8
	
	\bibitem[{{Woods} \& {Loeb}(1995)}]{Woods1995}
	{Woods} E., {Loeb} A., 1995, \apj, 453, 583
	
	\bibitem[{{Xu} \& {Zhang}(2017)}]{Xu2017}
	{Xu} S., {Zhang} B., 2017, \apjl, 846, L28
	
	\bibitem[{{Zhang} \& {Zhang}(2014)}]{zhangandzhang2014}
	{Zhang} B., {Zhang} B., 2014, \apj, 782, 92
	
\end{thebibliography}


%%%%%%%%%%%%%%%%%%%%%%%%%%%%%%%%%%%%%%%%%%%%%%%%%%

%%%%%%%%%%%%%%%%% APPENDICES %%%%%%%%%%%%%%%%%%%%%

\appendix
\numberwithin{equation}{section}
\section[A]{Planar Geometry}
\label{sec:planar}
In the main text, we considered emitting regions that are quasi-spherical in the co-moving frame. Here, we now consider the case of planar geometry. We assume that emission arises from material in a shell moving radially away from the explosion centre with a Lorentz factor $\Gamma_b$. Planar geometry implies the source has an angular size (in the co-moving frame) $\gtrsim R/\Gamma_b$ (where $R$ is the radial distance from the centre of explosion). The shell has a thickness $\Delta$ and is emitting while it is traversing between $R_0$ and $R_f$. We assume in what follows that the thickness is set by causality $\Delta \approx R_f/(2\Gamma_{\rm em}\Gamma_b)$ (see Appendix \ref{appendix:thickness}). Under these assumptions, the time it takes the shell to cross the acceleration region, the angular time-scale (the difference between arrival time of photons emitted at different angles towards the observer) and the observed variability time-scale are all directly related:
\beq
t_{\rm cr}=\frac{\Delta}{c}=\frac{R_f}{2c\Gamma_{\rm em}\Gamma_b}=t_{\rm ang}=t_{\rm v}
\eeq
The final time-scale in the problem is the expansion time-scale, the time it takes any portion of the bulk to traverse between $R_0$ and $R_f$. This time (converted to the observer frame) is smaller or equal to the former scales:
\beq
t_{\rm exp}=\frac{R_f-R_0}{2c\Gamma_{\rm em}\Gamma_b}\equiv \frac{\Delta R}{2c\Gamma_{\rm em}\Gamma_b}\leq t_{\rm v}.
\eeq
$t_{\rm exp}$ is also the heating and (assuming marginally fast cooling) the cooling time-scale. To summarize: $t_{\rm c}=t_{\rm h}=t_{\rm exp}\leq t_{\rm cr}=t_{\rm ang}=t_{\rm v}$.

We can now estimate the luminosity arising from the shell.	
In the co-moving frame, the maximum volume of particles heated during the acceleration front's crossing time is then $\Delta' \pi (R/\Gamma_b)^2$. In this frame, the emission lasts over a period $t_{\rm exp}'$. The observed luminosity is then at most
\beq
\label{eq:Lplanar}
L_{\rm obs}=\Gamma_{\rm em}^4 L'=\Gamma_{\rm em}^6 \frac{\epsilon_e}{\epsilon_B}\pi\frac{B_{\rm em}^2}{4\pi} t_{\rm v}^2 c^3\frac{R_f}{\Delta R}
\eeq
Once more, we take the magnetic field as implied by the condition of marginally fast cooling $B=B(\gamma=\gamma_s)$, but accounting now for the possibility that the heating/cooling are done on a time-scale that is shorter than $t_{\rm v}$ by a factor $R_f/\Delta R$. This yields an estimate of the Lorentz Factor
\beq
\Gamma_{\rm em}=8.4\times 10^3 \bigg(\frac{\epsilon_B}{\epsilon_e}\bigg)^{1/6}\bigg(\frac{\Delta R}{R_f}\bigg)^{1/2}\frac{\nu_{\rm p,5.5}^{1/3} L_{52}^{1/6}}{E_{\rm GeV}^{1/3}}.
\label{eq:Gammaplanar}
\eeq
This is the same as equation (\ref{eq:Gammaspher}), except for the dependence on $c/v_g$ and $\Delta R/R_f$. $\Delta R/R_f$ is limited by observations of pulse asymmetry in GRBs. Since with no relativistic motion in the bulk frame, $t_{\rm rise}/t_{\rm decay}=\Delta R/R_f$ \citep{BG2016}, observations of $t_{\rm rise}/t_{\rm decay}=0.3-0.5$ (e.g., \citealt{Nemiroff1994,Quilligan2002,Hakkila2011}) limit $\Delta R/R_f$ to the same ratio. Thus, once more we find that unless $E_{\rm GeV}$ is very large $\Gamma_{\rm em}\gg \Gamma_b$ is required.

Due to the planar geometry, the emission radius is now determined by equating the observed variability with the angular time-scale (as opposed to the similar expression in equation (\ref{eq:rspher}), which only provided a lower limit on the radius in the spherical case)
\beq
R=2 \Gamma_{\rm em} \Gamma_b c t_{\rm v}=3.4\times 10^{16} \frac{\Gamma_{b,2} \nu_{\rm p,5.5}^{1/3} L_{52}^{1/6}t_{\rm v,0.5}}{E_{\rm GeV}^{1/3} }\bigg(\frac{\epsilon_B}{\epsilon_e}\bigg)^{1/6}  \mbox{cm}
\label{eq:rplanar}
\eeq

A slightly different set-up, applicable for instance to internal shocks, is that instead of there being a shell passing through a fixed range of radii where heating occurs, there is an acceleration front (e.g. shock) that passes through the material with a velocity $v_g<c$ and triggers emission. This set-up is the 1D equivalent to the spherical geometry discussed in \S \ref{sec:luminosity}. The only difference from the case discussed above is that $t_{\rm cr}$ is increased by $c/v_g$. This ends up increasing the value of $\Gamma_{\rm em}$ by $(c/v_g)^{1/3}>1$ as compared with equation (\ref{eq:Lplanar}), and makes the argument for relativistic motion in the co-moving frame even more restrictive. Note that since, for shocks, we expect $v_g=c/\sqrt{3}$, the quantitative difference in the $\Gamma_{\rm em}$ limits between this case and the previous one, is rather small.

\section[B]{Thickness of emitting shell in planar geometry}
\label{appendix:thickness}
In Appendix \ref{sec:planar} we argue that if a GRB pulse arises from a shell of thickness $\Delta$, then $\Delta\approx R_f/(2\Gamma_{\rm em}\Gamma_b)$. We discuss here briefly the reasoning for this association. Consider first the case of $\Delta\ll R_f/(2\Gamma_{\rm em}\Gamma_b)$. In this case, the duration of a pulse will be set by $t_{\rm ang}\gg t_{\rm cr}$. However, GRB observations imply that the time between pulses is roughly equal to the pulses' durations \citep{NP2002}. This would mean that the distance between two consecutive shells, $\delta$ should be significantly different than their typical thickness ($\Delta$). Furthermore, this would require that $\delta/\Delta \approx t_{\rm ang}/t_{\rm cr}$, which seems contrived.

Consider now the situation $\Delta\gg R_f/(2\Gamma_{\rm em}\Gamma_b)$.
Note that by definition if the emission properties are highly variable throughout the shell, this would observationally be seen as multiple pulses (so that the original shell is in essence broken down to multiple shells). Thus, by construction, the emission is approximately constant over $\Delta$.
Therefore, $\Delta\gg R_f/(2\Gamma_{\rm em}\Gamma_b)$, would lead to very flat GRB pulses with $t_{\rm v}\approx t_{\rm cr}\gg t_{\rm decay}\approx t_{\rm ang}$, and similarly $t_{\rm v}\gg t_{\rm rise}$ in contrast with observations (see \cite{BG2016} for details).

%%%%%%%%%%%%%%%%%%%%%%%%%%%%%%%%%%%%%%%%%%%%%%%%%%

% Don't change these lines
\bsp	% typesetting comment
\label{lastpage}
\end{document}